\documentclass[twocolumn]{revtex4-1}
\usepackage{ifthen}
\usepackage{color}
\usepackage{graphics,graphicx,epsfig}
\usepackage{epsf,epstopdf,wrapfig}
\usepackage[export]{adjustbox}
\include{epsf}
\usepackage{enumitem}
\usepackage{amsmath,amssymb,amsfonts}
\usepackage{mathtools,amscd,wasysym,stackrel,textcomp,graphicx,color}
\usepackage[usenames,dvipsnames]{xcolor}
\usepackage[hidelinks]{hyperref}
	
\newcommand\mydots{\hbox to 0.8em{.\hss.\hss.}}
\newcommand{\beqn}{\begin{eqnarray}}
\newcommand{\eeqn}{\end{eqnarray}}
\newcommand{\beq}{\begin{equation}}
\newcommand{\eeq}{\end{equation}}

\newcommand{\abs}[1]{|#1|}

\newcommand{\be}{\begin{equation}}
\newcommand{\ee}{\end{equation}}

\newcommand{\commento}[1]{}

\newcommand{\comment}[1]{}

\newcommand{\sigmahat}{\hat{\sigma}}

\definecolor{ormar}{rgb}{.8,.2,0}

\newcommand{\mytitle}{Dissipation in non-steady state regulatory circuits}

\begin{document}

\title{\mytitle}

\author{P.~Szyma\'nska-Ro\.zek$^{1}$, D. Villamaina$^{2,3}$, J.~Mi\c{e}kisz$^{1}$, A.~M.~Walczak$^{3}$}

\affiliation{
	\normalsize{$^{1}$Faculty of Mathematics, Informatics, and Mechanics, University of Warsaw, Poland}\\
\normalsize{$^{3}$ Capital Fund Management, 23 rue de l'Universit\'e 75007 Paris, France}\\
	          \normalsize{$^{2}$ Laboratoire de physique de l'\'Ecole normale sup\'erieure (PSL University), CNRS, Sorbonne Universit\'e, and Universit\'e de Paris,  France}\\
	}
	
\date{\today}

\begin{abstract}
In order to respond to environmental signals, cells often use small molecular circuits to transmit information about their surroundings. Recently, motivated by concrete examples in signaling and gene regulation, a body of work has focused on the properties of circuits that function out of equilibrium and dissipate energy. We briefly review the probabilistic measures of information and dissipation and  use simple models to discuss and illustrate trade-offs between information and dissipation in biological circuits.  We find that circuits with non-steady state initial conditions can transmit more information at small readout delays than steady state circuits. The dissipative cost of this additional information proves marginal compared to the steady state dissipation. Feedback does not significantly increase the transmitted information for out of steady state circuits but does decrease dissipative costs. Lastly, we discuss the case of bursty gene regulatory circuits that even in the fast switching limit function out of equilibrium.

\end{abstract}

\maketitle

\section{Introduction}

Cells rely on molecular signals to inform themselves about their surroundings and their own internal state~\cite{Bialek_book}. These signals can describe the surrounding sugar type and concentration, which is the case of many bacterial operons, such as those used for lactose or galactose breakdown~\cite{Alon2006}. Signaling and activation of phosphorylated receptors provides a means of informing bacterial cells on faster timescales about a wide range of conditions including crowding, growth signals and stress~\cite{Phillips_book}. Triggered by these signals cells activate regulatory networks and cascades that allow them to respond in an appropriate way to existing signals. 

A response is usually caused by a change in the environment, which perturbs the previous state of the cell and the regulatory system. Specifically, if the regulatory circuit was functioning in steady state, a change in the concentration of the signaling molecule, or the appearance of a new molecule will kick it out of steady state. Here we investigate the response to such perturbations. 

The energy dissipated in a regulatory network comes on one hand from the fact that certain steps, for example producing proteins, require ATP. However, energy dissipation also measures how far out of equilibrium a given circuit functions by identifying irreversible (so ATP consuming) reactions~\cite{Lan2012, Mehta2012, Cao2015}. 

Regulatory circuits that function out of equilibrium (do not obey detailed balanced) dissipate energy, even if they produce the same amount of proteins as circuits that function in equilibrium. We are interested in exploring the constraints that energy dissipation imposes on circuit architectures. The motivation is not because of limiting energetic resources in cells; ATP is typically abundant~\cite{MiloPhillips, Moran2010} or can be generated by burning carbon present in the cell. Rather we consider energy dissipation as a measure of irreversibility that allows us to compare the irreversibility of signaling encoded in given circuit architectures.

In order to concentrate on this specific problem of dissipation coming from regulatory logic, we choose to study a simplified model with two binary elements: a receptor and a protein. Each element can be in one of two states: active or inactive, and its state regulates the state of the other element. The first element -- the receptor -- is our input that responds to changes in the environment, and the second element -- the regulatory protein such as a  kinase in a two component signaling cascade - is the output of our regulatory system. We do not take into account the ATP-ADP balance for these reactions, but concentrate on the dissipation coming from the regulatory computation. Effectively, we assume that while ATP is certainly needed, it is part of the hardware of the network and cannot be modified a lot. In turn, we are interested in the question of given a certain set of hardware, what is the best regulatory logic (software) we can implement.
 
Dissipation in molecular regulatory networks has received a lot of theoretical attention~\cite{Lan2012, Mehta2012, Cao2015, Seifert2012, Still2012, Ouldridge2017,tenwolde2016, itosagawa2015}. This line of thought goes back to the non-equilibrium scheme of kinetic proofreading~\cite{Hopfield1974, Ninio1975} in which energy is used for error correction of the signal. A more recent application~\cite{Lan2012} has shown that energy dissipation is also needed for regulatory circuits to adapt to external signals and respond accurately. A similar conclusion that energy dissipation is necessary was reached for molecular circuits that try to learn about external concentrations~\cite{Mehta2012} and  it was shown that the amount of dissipated energy limits reliable readout~\cite{Mehta2012, Barato2013, Barato2014, Bo2015, Govern2014,Ouldridge2017}. Results linking  information, dissipation and learning~\cite{Barrato2015, Brittain2017, Goldt2017} have been derived in the general framework of stochastic thermodynamics~\cite{Seifert2012, Parrando2015}. In the context of biochemical reactions, both continuous biochemical kinetics models~\cite{Mehta2012, Bo2015, Govern2014, Becker2013} and bipartite two state systems~\cite{Horowitz2014, Allahverdyan2009, Barato2014, Barrato2015, Sartori2014, Hartich2016} have been used in this context. Among other topics the link between dissipation and prediction has been explored, again showing that long term prediction requires energy expenditure~\cite{Becker2013, Still2012}, and the non-predictive part of the information about past fluctuations is linked to  dissipation~\cite{Still2012}. Most recently the links between information and dissipation have been studied in spatial systems~\cite{Falasco2018}.

A regulatory circuit fulfills a function and we assume that the goal of our network is to maximally transmit information between the input and output~\cite{Tkavcik2008}. This objective function has been studied before theoretically, using both binary and more detailed models~\cite{ Tkacik2011,  Tkavcik2009, Walczak2010, Tkavcik2012, Mugler2009, Rieckh2014, Sokolowski2015}. Others have also optimized the rate of information transmission~\cite{Tostevin2009, Tostevin2010, Ronde2010, Ronde2012}. Information transmission in regulatory circuits has also been investigated experimentally in fly development~\cite{Gregor2007, Gregor2007a, Dubuis2013}, NF$\kappa$B signaling~\cite{Cheong2011}, calcium signaling~\cite{Pahle2008} and dynamical readouts were compared to static information transmission between the input and output of ERK, calcium and NF$\kappa$B signaling networks~\cite{Selimkhanov2014}. While it is an arbitrary choice of the objective function for a regulatory network, and many networks do not optimize information transmission, it is rather unlikely that a circuit aimed at sensing and responding to the environment does not transmit any information about the signal to the output. The choice allows us to perform concrete calculations and investigate the trade-off between information and dissipation which are both tied to the logic of the regulatory system. 

Here, inspired by receptor-ligand binding, we use a simple two state system to build intuition about the trade-offs in information transmission, dissipation and functioning out of steady state. In a pedagogical spirit we remind the reader of the notions of dissipation and review some of our previous results from work that studied the trade-offs between information transmission and dissipation for regulatory circuits~\cite{Mancini2013, Mancini2015}. A signal often perturbs the system out of steady state, to which it then relaxes back. In this paper we calculate the non-equilibrium dissipation for circuits that function out of steady state and maximally transmit information between the input and a potentially delayed output given constraints on dissipation. 

Lastly we include some comments on dissipation in simple gene regulatory circuits with bursty transcription~\cite{KeplerElston, Raj2005, Friedman2006a, Walczak2005b, Cai2006, Golding2011, Desponds2016}. We show how even a fast switching gene promoter need not be in equilibrium. Our goal is not to provide an exhaustive review of the field but to illustrate with simple examples some trade-offs that appear in these molecular circuits.

\section{Model}
We consider a system consisting of two discreet random variables $z_t$ and $x_t$, evolving in time $t$, which describe the input state and output state of the systems, respectively. For simplicity we assume that $x$ and $z$ can take only two values: $+$ (active state) and $-$ (inactive state). The input state corresponds to the presence or absence of a signaling molecule (or a high or low concentrations of a signaling molecule), whereas the output state is activation or not of a response pathway or regulator. The specific regulatory interactions between them will be defined later within the specfic studied model(s). At every time $t$, the system is in one of four possible states ($z_t,x_t$): $(-,-)$, $(-,+)$, $(+,-)$, or $(+,+)$. The master equation for the temporal evolution of the conditional probability distribution $p(z_t,x_t|z_0, x_0)$ of the system is:
\begin{equation}
\frac{\partial}{\partial t}p(z_t,x_t|z_0, x_0) = -\mathcal{L}p(z_t,x_t|z_0, x_0),
\end{equation}
where $\mathcal{L}$ is a $4\times 4$ matrix with transition rates between the four states. We will be interested in the joint probability $p(x_{t},z_{0})$, that is we will look at the output variable $x$ at time $t$ and the initial state of the input variable $z$: 
\begin{equation}
p(x_{t},z_{0})=\sum\limits_{x_0,z_t=\pm1}^{}p(z_t,x_t|z_0,x_0)\cdot p(x_0,z_0).
\end{equation}
This probability is needed in the computation of the central quantity we optimize: the time--delayed mutual information between the initial state of the input and the state of the output at $t$ (defined in section~\ref{defMI}). After marginalization over possible states of $z_0$ we will obtain  $p(x_{t})=\sum\limits_{z_0}^{}p(x_{t},z_{0})$, which in turn is indispensable for calculating the dissipation of the system defined in section~\ref{defMI_diss}.

We restrict our analysis to symmetric models, in which we do we do not distinguish between the $(-,-)$ and $(+,+)$ states, and, analogically, between the $(-,+)$ and $(+,-)$ states. The symmetry of the model allows us to write the  probability distribution at any time $t$ as $p(x_t,z_0)=\left(\frac{1+\mu_t}{4},\frac{1-\mu_t}{4},\frac{1-\mu_t}{4},\frac{1+\mu_t}{4}\right)$, assuming the initial probability distribution also assumes the same symmetry: $p(x_0,z_0)=p_0=\left(\frac{1+\mu_0}{4},\frac{1-\mu_0}{4},\frac{1-\mu_0}{4},\frac{1+\mu_0}{4}\right)$. For the models in which the initial distribution is the steady state one, $p^{\text{init}}=p(x_0,z_0)=p^{\text{ss}}$, which imposes a condition on $\mu_0$.

\begin{figure}
\includegraphics[scale=1]{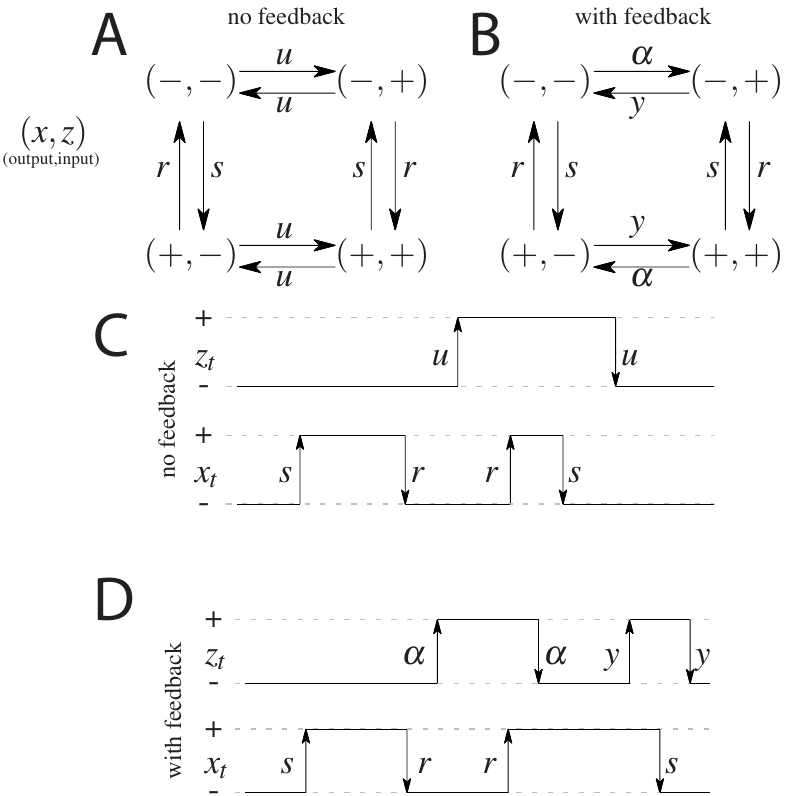}
\caption{\label{Fig1} A cartoon of the possible states and transitions  for both models: without feedback (A), and with feedback (B). Since there are two binary variables there are four states; transition rates are marked next to respective arrows. Note the symmetry between the ``pure'' ($(-,-)$ and $(+,+)$) states and the ``mixed'' states ($(-,+)$ and $(+,-)$) in both models. Representation of a possible time evolution of the system. Two variables flip between active ($+$) and inactive ($-$) states with respective rates. In the model without feedback (C) the output variable depends on the input variable (the output aligns to the input with rate $r$ or anti-aligns, with rate $s$), the input variable $z$ flips freely between its active and inactive state, regardless of the state of the output. In the model with feedback (D), there is a difference in rates of flipping of the input that depends on the state of the output.}
\end{figure}

\subsection{Model without Feedback}
The first, simplest model we analyze is a symmetric model in which only the input affects the output and there is no feedback from the output to the input. 
The output variable either aligns or anti-aligns to the input variable with rate $r$, regardless of the state of the input (see Fig.~\ref{Fig1}A).  The input variable $z$ flips between active and inactive states with rate $u$ and the output variable $x$ aligns with rate $r$ and anti-aligns with rate $s$ (see Fig.~\ref{Fig1}). The dynamics is given by a transition rate matrix given in Appendix~\ref{AppendixB}.

We calculate analytically the joint probability distribution $p(x_t,z_0)$ (a four-dimensional vector) and marginal probability distributions $p(x_t)$ and $p(z_0)$ (two-dimensional random vectors), needed to find the mutual information, that we will define in Eq.~\ref{defInfo},  as a function of the transition rates $u$, $s$, $r$, and a parameter $\mu_0$ that parametrizes the initial state of the system (see Appendix B). 
We set, without loss of generality, one rate equal to $1$, specifically $r=1$. The specific expressions for the probability distributions for the occupancy of the four states for the model without feedback are given in Appendix~\ref{AppendixB}. In steady state the probability distribution for the occupancy of the four states simplifies to $p^{\infty}=\left(\frac{u+1}{2 s+4 u+2},\frac{s+u}{2 s+4 u+2},\frac{s+u}{2 s+4 u+2},\frac{u+1}{2 s+4 u+2}\right)$.

\subsection{Model with Feedback}

In the second analyzed model we allow the input variable to be dependent on the output, i.e., we allow for a feedback from $x$ to $z$. We keep as much symmetry as possible, while still not distinguishing between the states $(-,-)$ and $(+,+)$, and between $(-,+)$ and $(+,-)$. The scheme is given in Fig.~\ref{Fig1}B. In terms of the rates we allow the original input  $z_t$ switching parameters, to be different depending on the state of the output $x_t$ introducing the rate $\alpha$ for anti-aligning the two variables and $y$ for aligning the two variables. The notion of input and output is no longer meaningful since both variables influence each other. We note that this scheme is not the most general model possible since we impose the symmetry between the 'pure' states, i.e., $(-,-)$ and $(+,+)$, and the 'mixed' states, i.e., $(-, +)$ and $(+,-)$, which reduces the number of parameters from 8 (as was studied in Mancini et al~\cite{Mancini2013}) to 4 (as was considered in  Mancini et al~\cite{Mancini2015}). The transition matrix for this model, and the steady state probabilities are given in Appendix~\ref{AppendixC}.

Consideration of the initial distribution multiplies the number of models. So far we have introduced two models - simple regulation and one with feedback. However, within both of them, we can either fix the initial distribution or let it be any four-dimension probability vector satisfying the symmetry condition (i.e., we let $\mu_0$ be any number between $-1$ and $1$). We will use the following notation:
\begin{itemize}
\item  $S$ - no feedback, stationary initial condition;
\item $\tilde{S}$ - no feedback, optimal initial condition;
\item $F$ - with feedback, stationary initial condition;
\item $\tilde{F}$ - with feedback, optimal initial condition.
\end{itemize}

\section{Information}\label{defMI}
The mutual information measured between the input $z$ at time $0$ and output $x$ at time $t$ is defined as~\cite{Cover1991, Tkacik2011}:
\begin{equation}\label{defInfo}
I[x_{t},z_{0}]=\sum_{x_{t},z_{0}}p(x_{t},z_{0})\log\frac{p(x_{t},z_{0})}{p(x_{t})p(z_{0})}.
\end{equation}
In order to analyse the system in its natural timescale, we set $t=\tau/\lambda$, where $\lambda$ is the inverse of the relaxation time (smallest, non-zero eigenvalue of the matrix $\mathcal{L}$) and calculate $I[x_{\tau}; z_0]=I[x_{\lambda\cdot t}; z_0]$. 

Again exploiting the symmetry of the problem, the mutual information can be written as
\beq
I[x_t, z_0] = \frac{1}{2}\left( (1 + \mu) \log(1 + \mu) + (1 - \mu) \log(1 - \mu)\right),
\eeq
where $\abs{\mu}\leq1$. Since we have fixed $r=1$, the symmetry of clockwise and counter-clockwise rotations is broken and $\mu \in [0, 1]$. Information is an increasing function of $\mu$ and is maximized at $I[x_t, z_0] =1$ bit for $\mu=1$. The specific values for $\mu$ are given in Appendix~\ref{AppendixB} and \ref{AppendixC} for the models with and without feedback.

 \section{Non-equilibrium Dissipation}\label{defMI_diss}

We consider the limitations on the regulatory architecture coming from having a fixed amount of energy to dissipate during the signaling process that transmits information. Large amounts of dissipated energy allow systems to function far out of equilibrium, whereas no dissipated energy corresponds to equlibrium circuits. We quantify the degree to which the system functions out of equilibrium by comparing the probability of a forward, $P_{\rightarrow }(\vec{x})$,  and backward, $P_{\leftarrow }(\vec{\tilde{x}})$, trajectory along the same path \cite{Crooks1998, Seifert2012}:
\beq
\label{initdiss}
\sigma= \sum_{\vec{x}} P_{\rightarrow }(\vec{x}) \log \frac{P_{\rightarrow }(\vec{x})}{P_{\leftarrow }(\vec{\tilde{x}})},
\eeq 
where the paths are defined as $\vec{x}=(x_1, x_2, \dots , x_N)$ and $\vec{\tilde{x}}= (x_{\text{N}}, x_{\text{N}-1}, \dots , x_1)$ and each state $x_i$ is a four dimensional probability of the input and output at time $i$.  Using the Markov nature of the transitions $P(x_{t+1}|x_t)$ we write the probability of the forward path starting from the initial state $x_1$ as
\beq
P_{\rightarrow} (\vec{x}) =P_{1}(x_1) \prod_{t=1}^{\text{N}-1} P_{t \rightarrow t+1}(x_{t+1}|x_t),
\eeq
and analogously for the backward path. Eq.~\ref{initdiss} now becomes:
\beqn
\sigma&=& \sum_{\vec{x}} P_{\rightarrow }(x_1, ..., x_\text{N}) \log \frac{P_{1}(x_1) \prod_{t=1}^{\text{N}-1} P_{t \rightarrow t+1}(x_{t+1}|x_t)}{P_{N}(x_N) \prod_{t=1}^{\text{N-1}} P_{t+1\rightarrow t}(x_{t}|x_{t+1})} \\ \nonumber
&=&  \sum_{\vec{x}} P_{\rightarrow }(x_1, ..., x_\text{N}) \log \frac{  \prod_{t=1}^{\text{N-1}} P_{t \rightarrow t+1}(x_{t+1}|x_t) P_{t}(x_t)}{  \prod_{t=1}^{\text{N-1}} P_{t+1\rightarrow t}(x_{t}|x_{t+1})P_{t}(x_{t+1}) },
\eeqn
where we multiplied both the numerator and the denominator by the same product of probabilities $P(x_2)\cdot ... \cdot P(x_N)$. Simplifying further and marginalizing over the elements of $\vec{x}$ not equal to $x_t$ or $x_{t+1}$:
\beqn
\sigma&=
&\sum_{t=1}^{\text{N}-1} \sum_{\vec{x}} P_{\rightarrow }(x_1,...,x_\text{N})\log\frac{P_{t \rightarrow t+1}(x_{t+1}|x_t)P_{t}(x_t)}{P_{t+1\rightarrow t}(x_{t}|x_{t+1})P_{t}(x_{t+1}) } \\ \nonumber
&=&\sum_{t=1}^{\text{N}-1} P_{\rightarrow }(x_t, x_{t+1}) \log\frac{P_{t \rightarrow t+1}(x_{t+1}|x_t)P_{t }(x_t)}{P_{t+1\rightarrow t}(x_{t}|x_{t+1})P_{t}(x_{t+1}) }\\ \nonumber
&=&\sum_t \sigma(t),
\eeqn
which defines the time dependent dissipation production rate, $\sigma(t)$.

Noting that $P_{t \rightarrow t+1}(x_{t+1}|x_t)=P_{t+1\rightarrow t}(x_{t}|x_{t+1})= P(x_{t+1}=i|x_t=j) $ and by explicitly defining the transition rates:
\beq
P(x_{t+1}=i|x_t=j) = w_{ij} \delta t + (1-w_{ij} \delta t) \delta_{ij},
\eeq
and renaming $P_{t}(x_{t+1})=p_j(t)$ and $P_y(x_{t})=p_i(t)$ we obtain \cite{Crooks1998,Tome2012, Seifert2012}:
\beq
\label{finalsigma}
\sigma(t)=\sum\limits_{i,j}^{}w_{ij}p_{j}(t)\log\frac{w_{ij}p_{j}(t)}{w_{ji}p_{i}(t)},
\eeq
that in the limit of $t\rightarrow\infty$ results in the \textit{steady state entropy dissipation rate}:
\beq\label{DissSS}
\sigma^{\text{ss}}=\sum\limits_{i,j}^{} p^{\text{ss}}_{j}w_{ij}\log\frac{w_{ij}}{w_{ji}},
\eeq
where $ p^{\text{ss}}_{j}$ is the steady state probability distribution. We describe an alternative derivation of dissipation in Appendix~\ref{Appendix_entropyprid}.

Again, we rescale the time in the above quantities by setting $t=\tau/\lambda$ ($\lambda$ being is the inverse of the relaxation time):
\beq
\hat{\sigma}(\tau)=\frac{1}{\lambda}\sigma(\tau/\lambda), \qquad  \hat{\sigma}^{\text{ss}}=\frac{1}{\lambda}\sigma^{\text{ss}}.
\eeq

\section{Setup of the optimization}

\begin{figure}
\includegraphics[scale=1.3]{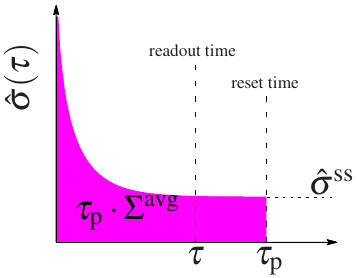}
\caption{\label{diss_cartoon}Schematic representation of system's relaxation. The entropy dissipation rate, $\hat{\sigma}(\tau)$ relaxes with time to its steady state value, $\hat{\sigma}^{\text{ss}}$. At $\tau_{\text{p}}$ the system is ``kicked out'' or reset, thus the pink area represents the total energy dissipated until that time. The information is collected at an earlier readout time $\tau$.}
\end{figure}

With these definitions we can ask what are the circuits that optimally transmit information given a limited constrained amount of steady state dissipation $\hat{\sigma}^{\text{ss}}$:
\beq
\label{optimization}
\rm{max}_{\mathcal{L}} \left[\mathcal{I}(\tau)-\sigmahat^{ss} \right],
\eeq
over the circuit's reaction rates, $\mathcal{L}$. The energy expense of a circuit that remains in steady state is well defined by this quantity. However the total expense of circuits that function out of steady state must be calculated as the integral of the entropy dissipation rate in Eq.~\ref{finalsigma} over the entire time the circuit is active, $\tau_p$, such as the duration of the cell cycle or the interval between new inputs that kick the system into the initial non-equilibrium state. After some time the circuit will relax to equilibrium (see the diagram in Fig.~\ref{diss_cartoon}) and its energetic expense is well described by the steady state dissipation. But the initial non-equilibrium steady state costs the system some energy.  We can compare the performance of circuits with different regulatory designs by considering the \textit{average energy expenditure} until a given time $\tau_p$:
\begin{align}\label{defSigma}
\Sigma^{\text{avg}}(\tau_{p}) = \frac{1}{\tau_{p}}\int\limits_{0}^{\tau_{p}}\hat{\sigma}(\tau) d\tau. 
\end{align}

We can foresee that circuits that spend most of their time in steady state will have their expenditure dominated by $\sigma^{\text{ss}}$, whereas circuits that spend a lot of time relaxing to equilibrium will be dominated by the additional out of steady state dissipation cost $\Delta \Sigma=\Sigma^{\text{avg}} - \hat{\sigma}^{\text{ss}}$. When $\tau_p\rightarrow \infty$, all circuits spend most of their time in equilibrium and the average integral in (\ref{defSigma}) converges to $\hat{\sigma}(\tau)  \rightarrow \hat{\sigma}^{\text{ss}}$ as ${\tau \rightarrow \infty}$, so that the cost is dominated by the steady state dissipation.

Using the steady state distribution for model S and Eq.~\ref{DissSS} we can evaluate the non-rescaled steady state dissipation calculated for the model without feedback 
\beq~\label{ss_ent_modelA}
\sigma^{\text{ss}}(u,s)=\frac{(s-1)u\log_{2}(s)}{1+s+2u}.
\eeq 
If we impose a non-equlibrium state by setting $s\rightarrow 0$, the dissipation rescaled by the characteristic decay time (the lowest non-zero eigenvalue given by the minimum of the two non-zero eigenvalues $1+s$, and $2u$) tends to infinity 
\beq\label{rescsigma}
\hat{\sigma}^{\text{ss}}(u,s)=\sigma^{\text{ss}}/\lambda=\frac{(s-1)u\log_{2}(s)}{(1+s+2u)\cdot\min(1+s,2u)}\xrightarrow[s\rightarrow 0]{} \infty,
\eeq
as expected. We also verify numerically that even in a non-steady state system that is kept out of equilibrium (Eq.~\ref{finalsigma}) the rescaled dissipation (Eq.~\ref{rescsigma}) tends to infinity, $\hat{\sigma}=\infty$ as $s\rightarrow 0$, for all $\tau$, $\mu_{0}$ and $u$.

The steady state dissipation rescaled by the  smallest eigenvalue for models $F$ and $\tilde{F}$ is~\cite{Mancini2015}:
\beq
\label{eq:sigmahat_fb}
\hat{\sigma}^{\text{ss}}(\alpha,s,y) = \frac{2(\alpha-s y)}{A(A-\rho)}\log_2\left(\frac{\alpha}{s y}\right),
\eeq
where
\beqn
A &=& 1+s +y+\alpha, \label{eq:A}\\
\rho &=& \sqrt{(1+s +y+\alpha)^2-8(s y+\alpha)} \label{eq:rho}.
\eeqn

\section{Results}

The task is to find maximal mutual information between the input and the output, with or without constraints, for all model variants, (regulation with and without feedback; starting at steady state, or starting out of steady state) and compare their performance -- the amount of information transmitted and the energy dissipated. To build intuition we first summarize the results of the unconstrained optimization obtained by Mancini et al~\cite{Mancini2013}. Then, a constraint will be set on the steady state dissipation rate $\hat{\sigma}^{\text{ss}}$ as in Mancini et al~\cite{Mancini2015}. We extend the latter results to models $\tilde{S}$ and $\tilde{F}$ by performing the optimization also with respect to the initial distribution. Finally, to compare not only the information transmitted in the models, but also its cost, we will calculate the average dissipation of the models.

In all cases we are looking for the maximum mutual information between the input at time $0$ and the output at time $\tau$, in the
space of parameters ($u$, $s$ and $r$ for the model without feedback and $\alpha$, $y$, $s$ and $r$ for the model with feedback).  We can also treat the initial distribution (parametrized by a single parameter, $\mu_0$), as an additional constraint or set $\mu_0$ to be equal to $\mu_{0}^{\text{ss}}$, i.e.,  fix the initial distribution to be the steady state one. Optimizing with a constraint is looking for the maximum of the function not in the whole parameter space ($\mathbb{R}^{\mathbb{N}}_{+}$), but on the manifold given by $\sigma^{\text{ss}}$(parameters) = constraint. Finally, to compare not only the information transmitted in the models, but also its cost, we will calculate the average dissipation of the models.

\subsection{Unconstrained optimization}

\begin{figure}
\begin{center}
\includegraphics[scale=1]{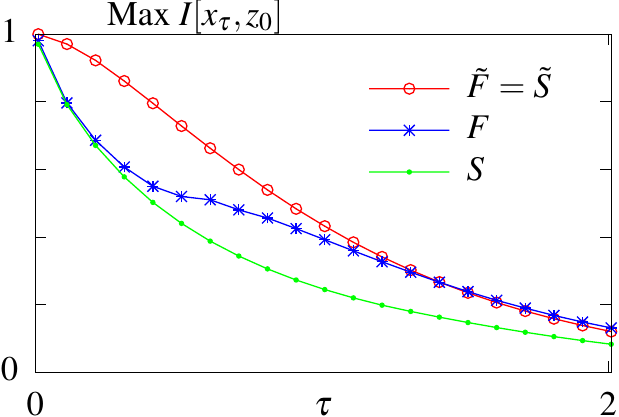}
\caption{\label{Fig6}Results of the unconstrained optimization - mutual information for the  models without feedback ($S$ and $\tilde{S}$) and with feedback ($F$ and $\tilde{F}$) with respect to the readout time $\tau$. Optimization done both when the initial distribution is fixed to its steady state value (no tilde) and when the parameter is subjected to optimization as well (with tilde).}
\end{center}
\end{figure}

The results of the unconstrained optimization are summarized in Fig.~\ref{Fig6}. As expected the maximum amount of information that can be transmitted decays with the readout time for all models. Feedback allows for better information transmission only in the case when the initial distribution is fixed to its steady state value. Optimizing over the initial distribution renders the models considered here without ($\tilde{F}$) and with feedback ($\tilde{S}$) equivalent. In this case the system relies on its initial condition and information loss is due to the system decorrelating and loosing information about its initial state. For a fixed initial distribution the model with feedback performs better than the model without feedback. We note that the feedback model considered here is a simplified model compared to the one studied in Mancini et al~\cite{Mancini2013}, with less parameters. A full asymmetric model with feedback can transmit more information than a model without feedback if the initial conditions are not in steady state. However these architerctures correspond to infinite dissipation solutions since all backward rates are forbidden and the circuit can never regain its initial state since one of the  states $i$ becomes absorbing, $p_{\infty}(y')=\delta_{y',i}$, and attracts the whole probability weight. We are therefore restricting our exploration of models with feedback to the subclass without an absorbing steady state.

The circuit architectures corresponding to the optimal solutions were discussed in previous work~\cite{Mancini2013, Mancini2015}. In short, the information-optimal steady state system uses rates that break the detailed balance and induce an order in visiting the four states $i$. Feedback increases the transmitted information for long time delays by implementing these cycling solutions  using a mixture of fast and slow rates.  Allowing for out of steady state initial conditions, circuits relax to absorbing final states that need to be externally reset. In this case the optimal solution with and without feedback is the same cycling architecture that simply relies on the decorrelation of the initial state.

\subsection{Constraining $\hat{\sigma}^{\text{ss}}$}

\begin{figure}
\begin{center}
\includegraphics[scale=1]{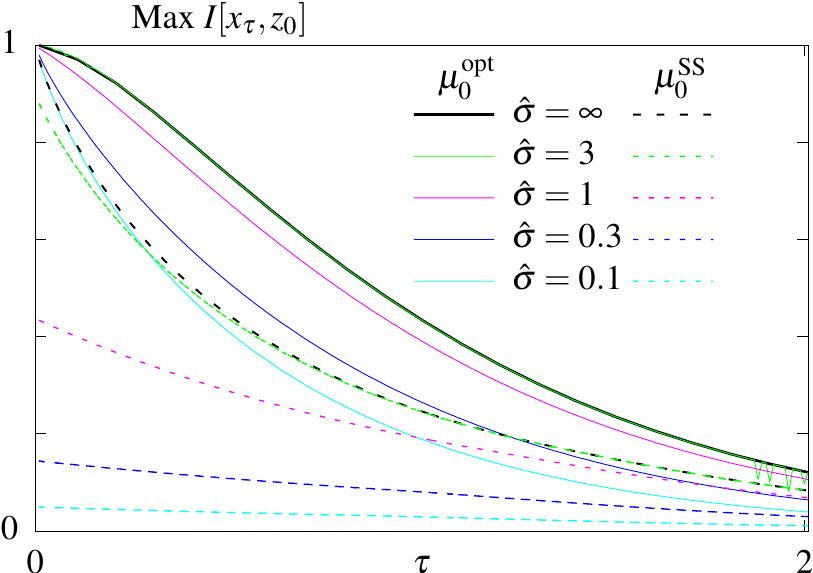}
\caption{\label{Fig7}Results of the optimization problem with constrained steady state dissipation for models without feedback. Optimal mutual information as function of the readout time, $\tau$, for different constrained steady state dissipation rates, $\hat{\sigma}^{\text{ss}}$, for the model $S$ (dashed lines) an $\tilde{S}$ (solid lines).}
\end{center}
\end{figure}

We next looked  for rates that maximize the transmitted information 
$I[x_{\tau},z_0]$ at a fixed time $\tau$ given a fixed steady state dissipation rate $\hat{\sigma}^{\text{ss}}$. We first plot the maximal mutual information as function of the readout time, $\tau$, for  models without feedback, $S$ (dashed lines) and $\tilde{S}$ (solid lines), (Fig.~\ref{Fig7}). Not surprisingly, maximum information is a decreasing function of $\tau$ for both models, larger values of steady state dissipation, $\hat{\sigma}^{\text{ss}}$, allow for more information transmitted, and model $\tilde{S}$ with optimized initial conditions transmits more information than model $S$, which remains in steady state. 

\begin{figure}
\begin{center}
\includegraphics[scale=0.8]{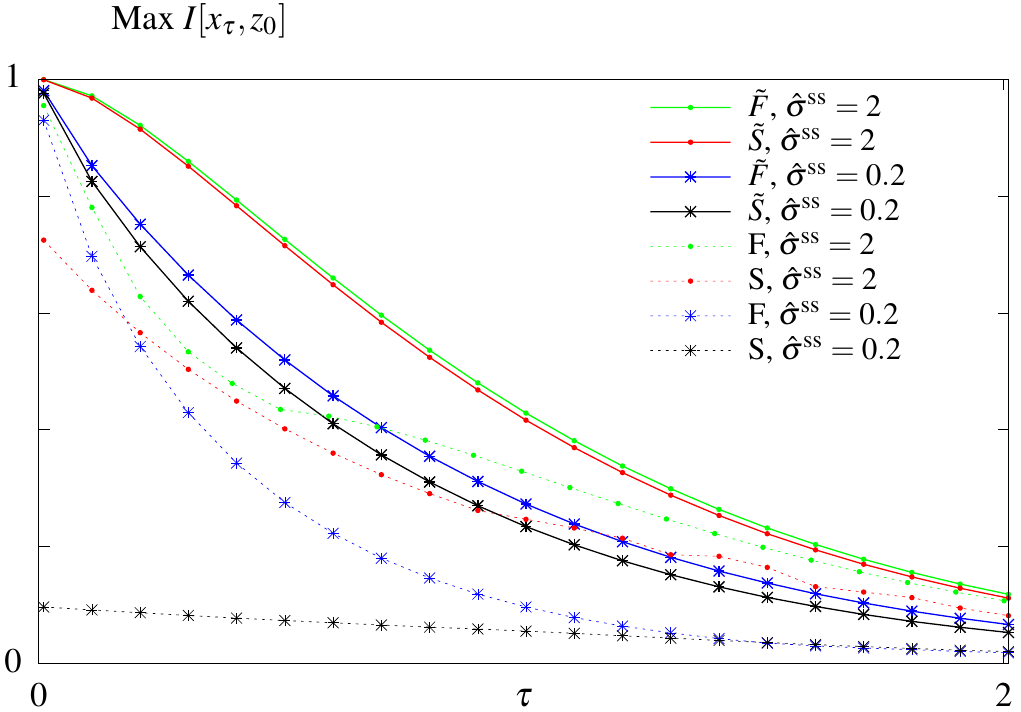}
\caption{\label{Fig9}Results of the optimization problem with constrained steady state dissipation for all four models. Optimal mutual information as function of the readout time, $\tau$, for two different  constrainedsteady state dissipation rates, $\hat{\sigma}^{\text{ss}}$, for the models $S$ and $F$ (dashed lines), and the models $\tilde{S}$ and $\tilde{F}$ (solid lines).}
\end{center}
\end{figure}

However comparing all four models, the conclusion about the  equivalence of the out of steady state model with ($\tilde{F}$) and without ($\tilde{S}$) feedback no longer holds when we constrain $\hat{\sigma}^{\text{ss}}$ (Fig.~\ref{Fig9}). The difference between optimal mutual information transmitted in models $\tilde{S}$ and $\tilde{F}$ is higher for systems that have smaller dissipation budgets $\hat{\sigma}^{\text{ss}}$, and, as shown previously (Fig.~\ref{Fig6}), the difference vanishes as  $\hat{\sigma}^{\text{ss}}\rightarrow\infty$. The remaining conclusions from Fig.~\ref{Fig7} hold:  models with feedback transmits more information than models without feedback and models with free initial distributions transmit more information than the steady state models, as in the unconstrained optimization case (Fig.~\ref{Fig6}). 

Phase diagrams describing the optimal architectures for steady state circuits are reported in Mancini et al~\cite{Mancini2015}. At large dissipation rates, the optimal out-of-equilibrium architectures exploit the increased decorrelation time of the system since cycling solutions are permitted. Close to equilibrium,  circuits with no feedback cannot transmit a lot of information. Circuits with feedback use a combination of slow and fast rates to transmit information. The optimal close to equilibrium architecture rapidly aligns the two variables $z_t$ and $x_t$ ($y>\alpha$, $s$ small), and slowly anti-aligns them, increasing the probability to be in the aligned $(+,+)$ and $(-,-)$ states. This results in a positive feedback loop. The same strategy of adjusting rates is used far from equilibrium but this time results in a cycling solution which translated into a negative feedback loop ($\alpha>y$, $s \approx 0 $). 

\begin{figure}
\begin{center}
\includegraphics[scale=1]{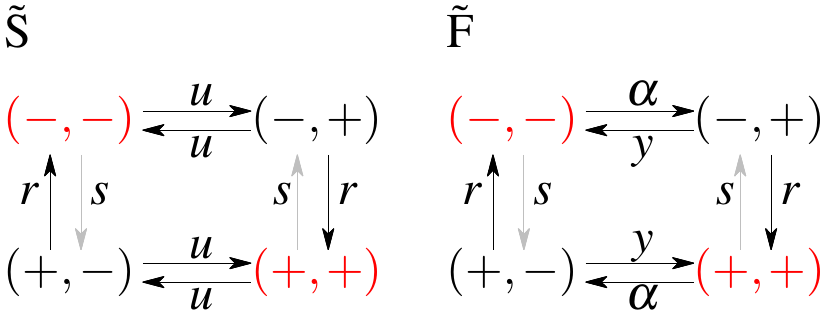}
\caption{\label{Fig_topo}A graphical representation of the optimal  circuits without ( $\tilde{S}$) and with ( $\tilde{F}$) feedback for delayed information transmission with optimized non-steady state initial conditions  with a constraint on steady state dissipation $\hat{\sigma}^{\text{ss}}$. The exact rate values depend on the value of $\hat{\sigma}^{\text{ss}}$ and examples are shown in Fig~\ref{SIFIG_modelA} (model $\tilde{S}$) and Fig~\ref{SIFIG_modelC} (model $\tilde{F}$). The depicted circuits are close to equilibrium. The gray arrow indicates a smaller rate than the black arrow. The red boxes show the optimized non-steady state initial states that have highest probability.}
\end{center}
\end{figure}

Allowing the circuit to function out of steady state optimizes the initial condition $\mu_0$ to be as far as possible from the equilibrium state. The optimal initial condition is $\mu_0=1$, where only the aligned states are occupied (the initial distribution is $p_0=(0.5, 0, 0, 0.5)$). This initial condition combined with $u<r$ and $s<r$ (Fig.~\ref{SIFIG_modelA}) decreases the decorrelation time and even a circuit with no feedback can transmit non-zero information. The rates of the circuits without feedback are simply set by the dissipation constraint, with $s\rightarrow 0$ for large dissipation and taking the value to balance $u$ close to equilibrium (Fig~\ref{SIFIG_modelA}). Optimal architectures for circuits far from equilibrium were reported in Mancini et al~\cite{Mancini2015} and close to equilibrium are shown in Fig.~\ref{Fig_topo}. Circuits with feedback also mostly rely on the decorrelation of the initial state. Since the majority of the initial probability weight is in the aligned states, the $y$ and $\alpha$ are always roughly equal (Fig~\ref{SIFIG_modelC}). Only at intermediate dissipation rates, $y$ slightly smaller than $\alpha$ and small $s$ stabliize the initial aligned states and further decrease the decorrelation time (Fig~\ref{Fig_topo}), encoding small negative feedback in the circuit. 

To summarize, for all $\hat{\sigma}^{\text{ss}}<\infty$, as well as for circuits that have no constraints on $\hat{\sigma}^{\text{ss}}$, we found $I(S)<I(\tilde{S)}$, $I(F)<I(\tilde{F})$, and $I(S)<I(F)$. Also, for all  $\hat{\sigma}^{\text{ss}}<\infty$, $I(\tilde{S})<I(\tilde{F})$, with $I(\tilde{S})\xrightarrow[\hat{\sigma}^{\text{ss}}\rightarrow\infty]{\text{}}I(\tilde{F})$, where we have defined the optimal mutual information $I(M)$ of a model $M\in\{S,\tilde{S},F,\tilde{F}\}$.

\subsection{Cost of optimal information}

\begin{figure*}
\begin{center}
\includegraphics[scale=1]{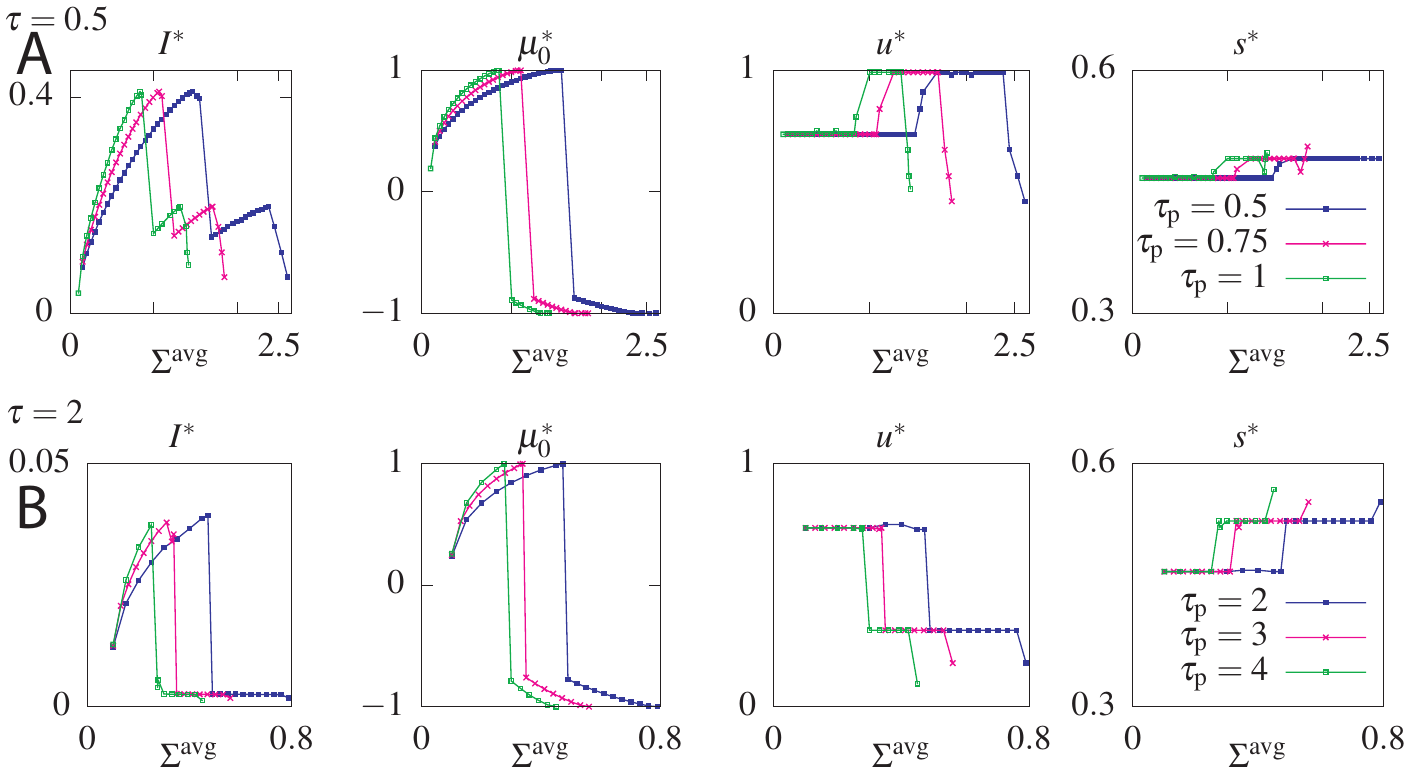}
\caption{\label{Fig10}Optimal mutual information (I*) and optimal parameters $\mu_0$, $u$, and $s$ for the $\tilde{S}$ model without feedback as function of the average dissipation, $\Sigma^{\text{avg}}$, for two values of the readout time, $\tau=0.5$ (A panels), and $\tau=2$ (B panels), and three values of the reset time, $\tau_p$ (different colours of curves). Steady state dissipation, $\hat{\sigma}^{\text{ss}}$, was fixed to $0.1$.}
\end{center}
\end{figure*}

The maximum information is  obtained for maximum {\it allowed} steady state dissipation. Interestingly the steady state dissipation $\hat{\sigma}^{\text{ss}}$ combined with the circuit topology impose a constraint on the   maximum allowed $\Sigma^{\text{avg}}(\tau_p)$. This result follows from the fact that the system strongly relies on the initial condition to increase the information transmitted at small times. Larger $\mu_0$ values allow the system to transmit more information, since the equilibration time is longer. However, fixing the value of $\hat{\sigma}^{\text{ss}}$ constrains the allowed value of $\mu_0$ that determine the initial condition. To gain intuition, additionally to fixing $\hat{\sigma}^{\text{ss}}$, we will fix the mean dissipation  $\Sigma^{\text{avg}}(\tau_p)$ until a reset time $\tau_p>\tau$ and find the transition rates returning the optimal mutual information for a chosen readout time $\tau\leq\tau_p$. The results of this optimization presented in Fig.~\ref{Fig10}, show that as $\Sigma^{\text{avg}}$ increases, $\mu_0$ tends towards $1$, which corresponds to a probability distribution where only the asymmetric states ($p_0=(0.5, 0, 0, 0.5)$) are occupied and the transmitted information increases. Further increasing dissipation shows that the $\hat{\sigma}^{\text{ss}}$  constraint can be satisfied in two ways: either by a positive or negative $\mu_0$. Not only does the positive $\mu_0$ transmit more information but the negative $\mu_0$ is forbidden by our choice of $r=1$. Above a certain value of  $\hat{\sigma}^{\text{ss}}$ only the forbidden negative $\mu_0=-1$ branch corresponding to an  initial distribution with all the weight in the anti-aligned states $p_0=(0,0.5,0.5,0)$ remains (if we chosen the counter clockwise  solutions by fixing $s=1$ this probability vector would have been the maximally informative initial state). The system cannot fulfill the constraint of such high dissipation.  If we do not constrain $\hat{\sigma}^{\text{ss}}$ we find that the maximum information corresponds to $\mu_0=1$~\cite{Mancini2013}, which we report in our analysis below. 

\begin{figure}
\begin{center}
\includegraphics[scale=0.9]{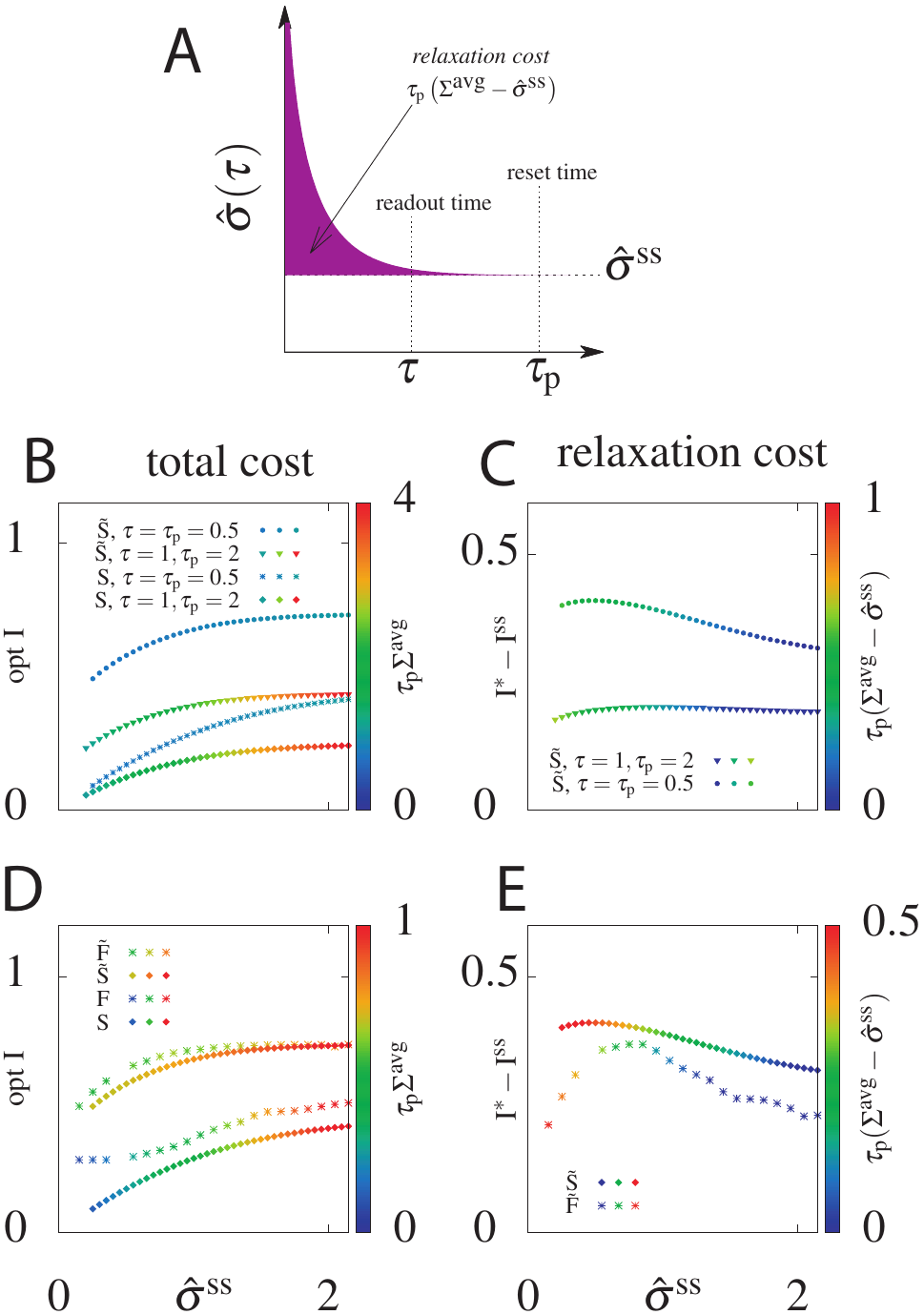}
\caption{\label{Fig13} (A) Cartoon depicting the relaxation cost (pink area) $\tau_p (\Sigma^{\text{avg}}-\hat{\sigma}^{\text{ss}})$ of the system equilibrating from a non-steady state initial state, and thus $\hat{\sigma}(\tau)\neq\hat{\sigma}^{\text{ss}}$. (B) The total cost, $\tau_p\Sigma^{\text{avg}}$, of the optimal information transmitted as a function of the steady state entropy dissipation rate, $\tau_p\hat{\sigma}^{\text{ss}}$, for models without feedback, that start with the steady state distribution, $S$, and that optimize the initial distribution, $\tilde{S}$. Results shown for two choices of reset $\tau_p$ and readout $\tau$ timescales. For the steady state models $\tau_p\Sigma^{\text{avg}} = \tau_p\hat{\sigma}^{\text{ss}}$.  (C) The information gain, $I^{*}-I^{\text{ss}}$,  of the optimized initital condition model ($\tilde{S}$) compared to the steady state initial condition model ($S$) and the relaxation cost, $\tau_p (\Sigma^{\text{avg}-\hat{\sigma}^{\text{ss}}})$, as a function of  the steady state entropy dissipation rate for the same choices of $\tau_p$ and $\tau$ as in panel (B). (D) Comparison of the optimal delayed information and total dissipative cost as a function of the steady state entropy dissipation rate for all four models: without feedback ($S$, $\tilde{S}$) and with feedback ($F$, $\tilde{F}$), with the initial distribution equal to the steady state one ($S$, $F$) or optimized over ($\tilde{S}$, $\tilde{F}$). $\tau=\tau_p=0.5$. (E) The information gain and relaxation cost of circuits with optimized initial conditions compared to steady state ones for the models with ($\tilde{F}$) and without feedback ($\tilde{S}$). $\tau=\tau_p=0.5$. }
\end{center}
\end{figure}

We have seen that the for both models, if we can choose the initial distribution, instead of starting from the steady state, we can significantly increase the transmitted information. What is the "cost" of this choice of initial distribution? To estimate this total cost we calculate the average dissipation during time $\tau_p>\tau$, $\tau_p\Sigma^{\text{avg}}(\tau_p)$, for the circuit with the highest mutual information attainable for a given steady state dissipation rate rate $\hat{\sigma}^{\text{ss}}$ if we allow the initial condition to be out of the steady state (Fig.~\ref{diss_cartoon}). We also introduce the {\it relaxation cost}, $\tau_p (\Sigma^{\text{avg}}-\hat{\sigma}^{\text{ss}})$ (Fig.~\ref{Fig13} A), as the additional energy dissipated above the steady state value. As argued already, the systems that starts at steady state, i.e., for which $\mu_0=\mu_0^{\text{ss}}$, will not pay an additional cost (see Fig.~\ref{diss_cartoon}, for $\mu_{0}=\mu_{0}^{\text{ss}}$ the function of $\hat{\sigma}(\tau_p)$ is constant, equal to  $\hat{\sigma}^{\text{ss}}$). In this case the mean total dissipation, $\Sigma^{\text{avg}}(\tau_p)$, will be equal to  $\hat{\sigma}^{\text{ss}}$ and the relaxation cost goes to zero.

As shown in Fig.~\ref{Fig13} B, the total cost (z-axis, in colour) generated was only slightly larger for $\tilde{S}$ than for $S$ and the difference is more pronounced only for relatively small $\hat{\sigma}^{\text{ss}}$, where the cost in the steady state circuits goes to zero. This result holds for different combinations of delay readout times $\tau$ and reset times $\tau_p$, although the value of the total cost naturally increases with $\tau_p$. As discussed above, more information can be transmitted at shorter times and by optimizing over the initial condition. 

In order to quantify the intuition that $\tilde{S}$ transmits more information than $S$ at a small price, we plotted in Fig.~\ref{Fig13} C the \textit{information gain}, $I^{*}-I^{\text{ss}}$, and the relaxation cost with respect to $\tau_p(\hat{\sigma}^{\text{ss}})$. $I^{*}-I^{\text{ss}}$ is the difference between the optimal information when the initial distribution is free to be optimized over ($\tilde{S}$) and the optimal information for the system with a steady state initial distribution ($S$). It quantifies the additional cost from optimizing the initial condition of the gain in information transmission. The relaxation cost is almost the same regardless of the reset time, $\tau_p$. The relaxation cost and the information gain decrease with increasing steady state dissipation, $\hat{\sigma}^{\text{ss}}$, as in this regime even the steady state system is able to have slow decorrelation by tuning the switching rates.

This analysis shows that higher optimal mutual information obtained by optimizing over the initial distribution does not generate significantly higher costs. The same result holds when comparing models with feedback $F$ and $\tilde{F}$ (Fig.~\ref{Fig13} D). The information increase from feedback in the $\tilde{F}$ model with optimized initial conditions compared to the $F$ steady state model is minimal at large $\hat{\sigma}^{\text{ss}}$ (as expected from Fig.~\ref{Fig9}). While the $\tilde{F}$ model with feedback always transmits more information than the $\tilde{S}$ model without feedback, the total average cost for all $\hat{\sigma}^{\text{ss}}$ is smaller for the $\tilde{F}$ model with feedback than  for the $\tilde{S}$ model without feedback. This results means that even when feedback does not  increase the transmitted information compared to models without feedback, it decreases the total cost. 

The information gain of circuits with optimized initial conditions compared to steady state circuits is larger for the $\tilde{S}$ model without feedback than the $\tilde{F}$ model with feedback (Fig.~\ref{Fig13} E) and the relaxation cost decreases monotonically with increasing $\hat{\sigma}^{\text{ss}}$. In both the case with and without feedback there is a non-zero and non-infinite value of steady state dissipation where the information gain from optimizing the initial condition is largest. In summary, optimizing the initial condition nearly always incurs a cost, however it absolutely always results in a significant information gain. Table~\ref{table} summarizes the comparison of the optimal transmitted information $I(M)$ and total cost $C(M)$ for all four models $M\in\{S,\tilde{S},F,\tilde{F}\}$.

\begin{table}[h!]
    \begin{center}
    \begin{tabular}{l|c|c}
    \rule[-2ex]{0pt}{5.5ex} & $I^{\text{\,opt}}$ & Cost \\ \hline
    \rule[-2ex]{0pt}{5.5ex} $S$, $F$ & $I(S)<I(F)$ &  $C(S)=C(F)$ \\ \hline
    \rule[-2ex]{0pt}{5.5ex} $\tilde{S}$, $\tilde{F}$ & $I(\tilde{S})\leq I(\tilde{F})$  & $C(\tilde{S})>C(\tilde{F})$ \\ 
    \end{tabular}
    \end{center}\caption{\label{table}Comparison between the four models, $S$, $F$, $\tilde{S}$, and $\tilde{F}$ in terms of optimal mutual information, $I^{\text{opt}}$, and the cost (value of $\Sigma^{\text{avg}}$ calculated with optimal rates), $C$.}
\end{table}

\section{Gene regulatory circuits}

The coupled two state system model considered above can be thought of as a simplified model of receptor--ligand binding. It can also be considered as an overly simplified model of gene regulation where the input variable describes the presence or absence of a transcription factor  and the output -- the activation state of the regulated gene. However, the continuous nature of transcription factor concentrations has proven important when considering information transmission in these systems~\cite{Tkavcik2009,  Tkacik2011}. We will not repeat the whole optimization problem for continuous variables but we calculate and discuss the form of dissipation in the simplest gene regulatory module that can function out of equilibrium. 

\subsection{Bursty gene regulation}\label{Master_eq}

The simplest gene regulatory system that can function out of equilibrium is a model that accounts for transcriptional bursts~\cite{KeplerElston, Raj2005, Friedman2006a, Walczak2005b, Cai2006, Golding2011, Desponds2016}. The promoter state has two possible states: a basal expression state where the gene is read out a basal rate $R_0$ and an activated expression state where the gene is read out at rate $R_1$. The promoter switches between these two states by binding a transcription factor present at concentration $c$, with rate $k_+$ and unbinds at a constant rate $k_-$. The probability that there are $g$  product proteins of this gene in the cell (we integrate out the mRNA state due to a separation of timescales) is $P(g)=P_0(g)+P_1(g)$, where $P_0(g)$ describes the probability that the promoter is in the basal state and there are $g$ proteins and $P_1(g)$ describes the analogous probability for the promoter to be in the activated state. The probability distribution evolves both due to binding and unbinding of the transcription factor and to protein production and degradation (with rate $\tau^{-1}$) according to
\begin{eqnarray}
\frac{d P_{0}(g)}{dt}&=& \frac{g+1}{\tau} P_{0}(g+1)+k_{_{-}} P_{1}(g)+R_{0} P_{0} (g-1)+\\ \nonumber
&&- \left(k_{_{+}}c +\frac{g}{\tau}+R_{0}\right)P_{0}(g),\label{first_eq}\\
\frac{d P_{1}(g)}{dt}&=& \frac{g+1}{\tau} P_{1}(g+1)+k_{_{+}}c P_{0}(g)+R_{1} P_{1} (g-1)+\\ \nonumber
&&- \left(k_{_{-}}+\frac{g}{\tau}+R_{1}\right) P_{1}(g).\label{second_eq}
\end{eqnarray}
These equations can be solved analytically in steady state in terms of special functions~\cite{Hornos2005,Szymanska2011}. In the limit of fast promoter switching ($k_{_{+}}$ and $k_{_{-}}$ go to infinity and their  ratio $K\equiv k_{_{+}}/k_{_{-}}$ is constant) the system is well described by a Poisson distribution 
\begin{equation}
P^{*}_{1}(g)= \frac{1}{1+c K} \frac{(R_{ef}\tau)^{g}}{g!}\textrm{exp}(-R_{ef}\tau) = c K P^{*}_{0}(g)\label{steady_fast_switch}
\end{equation}
where $R_{eff}$ is an effective production rate:
\begin{equation}
R_{eff}=\frac{k_{_{+}}cR_{1}+k_{_{-}}R_{0}}{k_{_{+}}c+k_{_{-}}}.
\end{equation}

The total steady state dissipation $\sigma^{ss}=\sigma_{0}+\sigma_{1}+\sigma_{2}$ calculated from Eq.~\ref{DissSS} can be split in three parts, where 
  \begin{eqnarray}
\sigma_0&=&\sum_{g} \left(P^{*}_0(g)k_{_{+}} c-P^{*}_{1}(g)k_{_{-}}\right) \textrm{log}\frac{k_{_{+}} c}{k_{_{-}}},\\
\sigma_1&=&\sum_{g} \left(P^{*}_0(g) R_{0} \textrm{log} (R_{0}\tau)+P^{*}_1(g) R_{1} \textrm{log} (R_{1}\tau)\right)\label{sigma1},\\
\sigma_2&=& -\sum_{g} P^{*}_0(g)\left[ R_{0} \textrm{log} (g+1)+\frac{g}{\tau}\textrm{log}\frac{R_{0}\tau}{g}\right]+\\ \nonumber
&&- \sum_{g}P^{*}_1(g) \left[R_{1} \textrm{log} (g+1)+\frac{g}{\tau}\textrm{log}\frac{R_{1}\tau}{g}\right]\label{sigma2}.
\end{eqnarray}
 The first two expressions can be simplified using the normalization relations $\sum_{g}\left(P^{*}_{0}(g)+P^{*}_{1}(g)\right)=1$ and $\sum_{g}P^{*}_{1}(g)=\frac{k_{_{+}}c}{k_{_{-}}+k_{_{+}}c}$ obtaining:
\begin{eqnarray}
\sigma_{0}&=&0\\
\sigma_{1}&=& \frac{1}{k_{_{-}}+k_{_{+}}c} \left( R_{0} \textrm{log}(R_{0}\tau) k_{_{-}}+R_{1} \textrm{log}(R_{1}\tau) k_{_{+}}c\right).
\end{eqnarray}
We now use these results to examine steady state dissipation in the equilibrium limit and the limit of the fast switching promoter. Similar results but in slightly different limits were obtained in Ref.~\cite{Mehta2012}.

{\bf Equilibrium Limit.} 
Equilibrium is surely achieved if there is only one promoter state. In terms of our model this corresponds to  $k_{+}$ is vanishing and $k_{-}\neq
0$. In this limit the activated state is never occupied and the steady
state probability goes to $P^{*}_{1}(g)\equiv 0$. Eqs.~\eqref{first_eq} and \eqref{second_eq} result in a  Poisson distribution with mean $R_{0}\tau$ and we can verify that detailed balance is satisfied
\begin{equation}
P^{*}_{0}(g) W_{(g\to g\pm 1)}=P^{*}_{0}(g\pm1) W_{( g\pm 1\to g)},
\end{equation}
as confirmed by $\sigma_{2}=-\sigma_{1}$ in Eqs.~(\ref{sigma1}-\ref{sigma2}).

{\bf Fast promoter switching limit.} 
In the fast promoter switching limit  the dissipation of the system is:
\begin{equation}
\sigma_{_{FS}}=\frac{c K}{(1+c K)^{2}} (R_{0}-R_{1})\textrm{Log}\left(\frac{R_{0}}{R_{1}}\right).\label{fast_diss}
\end{equation}
$\sigma_{_{FS}}$ is always positive, but the equilibrium regime is reached only if $k_{_{-}}$ or $k_{_{+}}$ asymptotically vanish. For finite binding and unbinding rates the system is not in equilibrium despite being well described by an equilibrium-like steady state probability distribution. Since this example is mainly presented as a pedagogical application of dissipation, for completeness we derive similar results in the Langevin description in Appendix~\ref{Langevin}, discussing the differences in dissipation arising from model coarse graining~\cite{Crisanti2012, Puglisi2015,Busiello2019}.

\section{Discussion}
All living organisms, even the most simple ones, in order to adapt to the environment, must read and process information. In the case of cells, transmitting information means sensing chemical stimuli via receptors  and activating biochemical pathways in response to these signals. Such reading and transmitting signals comes at a price - it consumes energy. There are plenty of possible architectures of these regulatory circuits, yet not all of them are found in nature~\cite{Alon2006}. The question arises why some network architectures are frequent and others non-existing. One way to approach such a question is to optimize a (specific) function by a choosing the circuit architecture - it could be for example noise (minimization)~\cite{saunders2009}, time-delay of response (minimization)~\cite{Alon2006} or information transmitted between the input and output (maximization)~\cite{Tkacik2012}.

Two different circuits can produce and use the same amount of proteins, but the energy dissipated in them is different. In other words, we assume that while ATP is certainly needed in a molecular circuit, it is part of the hardware of the network and cannot be modified a lot. Instead, we asked about the best regulatory logic (software) we can implement, given a certain set of hardware. For this reason we worked with a simplified binary representation of the circuits to concentrate on the regulatory computation.

Our main previous findings about steady state circuits can be related to tasks performed by the circuits~\cite{Mancini2015}.  Circuits that function close to equilibrium transmit information optimally using positive feedback loops that are characteristic of long-term readouts responsible for cell fate commitment~\cite{Xiong2003, Tanaka2008}. Circuits that function far from equilibrium transmit information using negative feedback loops that are representative of shock responses that are transient but need to be fast~\cite{Guisbert2004, Lahav2004}. Therefore cells may implement non-equilibrium solutions when fast responses are needed and rely on equilibrium responses when averaging is possible and there is no rush. This results agrees with the general finding of Lan et al~\cite{Lan2012} for continuous biochemical kinetics that negative feedback circuits always break detailed balance and such circuits function out of equilibrium.

In general in steady state we find that models with feedback  significantly outperform models without feedback in terms of optimal information transmission between the two variables, but the respective costs of optimal information are the same. Circuits close and far to equilibrium rely on a mixture of slow and fast timescales to delay relaxation and transmit information. The only other solution available in our simple setting is using the initial condition, which is efficient in terms of information transmission but costly. 

Here we identified two properties linked to feedback: it does not necessarily transmit more information if we are allowed to pick an optimal initial condition compared to a system without feedback. Yet in this case implementing a circuit with feedback can reduce the non-equilibrium costs. In general, introducing an optimized intitial condition incurs a cost, but this cost is often minimal, especially taking into account the information gained. This cost is interpretable biologically as the external energetic cost needed to place the system in a specific initial condition. This cost must be provided by the work of another regulatory element or circuit or an external agent or force. This specific initial condition requires poising the system in a specific point. Yet it does not seem biologically implausible, let alone impossible, to "prepare" the intitial state after cell division or mitosis, or upon entering a new phase of the cell cycle~\cite{Tyson2015}. For example, a specific gene expression state or receptor state (e.g. ($+,+$ or $-,-$)) seems easily attainable.

One could look at these results from two perspectives: on the one hand argue that circuits with feedback transmit more information in the steady state setting; on the other hand feedback exhibits frugality in expenses in the case of optimized initial distributions. One could also defend the models without feedback stating that they can be only slightly worse in terms of information transmission (optimized initial distribution case) and can be found to dissipate the same amount of energy (steady state initial distribution). All circuits will reach steady state, however especially during fast processes such as development~\cite{Lucas2018} or stress response~\cite{Lahav2004}, the information transmitted during short times may be what matters for downstream processes.  In general regardless of the timescale, circuits with feedback perform better (or equally well) than  regulatory system with no feedback, both in terms of information transmission and the cost of transmitting this optimal information.

The learning rate is another quantity that has been useful in studying bipartite systems in stochastic thermodynamics~\cite{Barrato2015, Brittain2017, Goldt2017}. The learning rate, defined as $l_x=\partial_{\tau} I[z_{\tau}, x_{t+\tau}]|_{\tau=0}$, gives the instantaneous increase in information that the output variable has by continuing to learn about the input variable. 
We calculate the learning rate for our informationally-optimal models when they are in steady state (Fig.~\ref{SIFIG_learningrate}). For models without feedback the learning rate is bounded by $\sigma_x$ (as defined in Appendix~\ref{learning_rate}), such that $\eta={\ell}_x/\sigma_x\leq 1$. It this case the interpretation of the learning rate allows us to estimate how closely the output variable is following the input variable and positive learning rates are indicative of adaptation and learning. Not surprisingly we find that the model with steady state initial conditions has a larger learning rate than the model with optimized initial conditions since model $\tilde{A}$ relies less on the parameters of the network than model $A$ to transmit information and more on the initial conditions (that are forgotten in the steady state calculation). Calculating a time delay dependent learning rate would be more informative. The learning rate also increases with $\hat{\sigma}$, in agreement with previous statements that learning is easier far from equilibrium~\cite{Barrato2015, Mehta2012, Lan2012}. We  also performed the same calculation for models with feedback but as was pointed out previously \cite{Ueda2012, Sagawa2012, Brittain2017}, the interpretation of the learning rate becomes less clear in these systems since input and output are no longer clearly defined. Instead the above one-sided definition should be replaced by a time integral over the trajectory to distinguish if the learning is  of the other variable ($z$) or a previous instance of the same variable ($x_{t-\tau}$). The calculated quantity instead tells us about the ability of $x$ to respond to $z$, assuming $z$ was fluctuating freely. In that sense a positive value of $l_x$ tells us that the dynamics of the two variables of the circuit are not completely decoupled in steady state, except in the case of model $F$ close to equilibrium. Our results tell us that equilibrium imposes a symmetry between input and output, which is broken either by initial conditions ($\tilde{F}$ at small $\hat{\sigma}$) or large dissipation. 

Lastly, for pedagogical purposes we attempted to discuss the link between dissipation calculations that are often performed on binary regulatory systems and continuous variables, showing that the simplest model of bursty transcription can result in non-zero dissipation, even in the fast switching limit where the steady state equilibrium Poisson distribution is recovered.  Bursty gene expression is wide spread from bacteria~\cite{Cai2006, Golding2011}, yeast~\cite{Raser2004} to invertebrates~\cite{Lucas2018, Desponds2016} and mammals~\cite{Raj2005}. Bursty self-activating genes in intermediate fast switching regimes have also been  shown to have different stability properties than pure equilibrium systems, due to non-equilibrium cycling through the coupled promoter and protein  states~\cite{Walczak2005a}. While cells are not energy limited, the discussion recounted in this paper may suggest that different modes of regulation (including burstiness) may be better suited for slow and fast responses.

{\bf Acknowledgements.} We thank T. Lipniacki, A. Nourmohammad and T.~Mora for helpful discussions. This work was in part supported by MCCIG no. 303561. J. M. would like to thank the National Science Centre (Poland) for financial support under Grant No. 2015/17/B/ ST1/00693.

\appendix

\section{Model without feedback}\label{AppendixB}
The transition matrix for the model without feedback reads:
\begin{eqnarray}\label{LA}
\mathcal{L}=\left(
\begin{array}{cccc}
u+s & -u  & -r  & 0\\
-u  & u+r & 0   & -s\\
-s  & 0   & u+r & -u\\
0   & -r  & -u  & u+s\\
\end{array}
\right),
\end{eqnarray}
where the rates are defined in Fig.~\ref{Fig1} A. By matrix diagonalization we find the eigenvalues and eigenvectors and calculate the probability distribution $p(x_{\tau}, z_0)$ at time $\tau$ for the four states, 

\begin{eqnarray}
&&p_{+,+}(\tau)=p_{-,-}(\tau)= \\ \nonumber
&&\frac{e^{-\frac{\tau  (s+2 u+1)}{\lambda }}}{4 (s-2 u+1)}\Big((\mu_0 (s-2 u+1)+s-1) e^{\frac{2 \tau  u}{\lambda }}+ \\ \nonumber
&& (s-1) \left(-e^{\frac{(s+1) \tau }{\lambda }}\right)+(s-2 u+1) e^{\frac{\tau  (s+2 u+1)}{\lambda }}\Big),
\end{eqnarray}

and

\begin{eqnarray}
&&p_{+,-}(\tau)=p_{-,+}(\tau)=\\ \nonumber
&&\frac{e^{-\frac{\tau  (s+2 u+1)}{\lambda }}}{4 (s-2 u+1)} \Big(-(\mu_0 (s-2 u+1)+s-1) e^{\frac{2 \tau  u}{\lambda }}+\\ \nonumber
&&(s-1) e^{\frac{(s+1) \tau }{\lambda }}+(s-2 u+1) e^{\frac{\tau  (s+2 u+1)}{\lambda }}\Big).
\end{eqnarray}

The steady state distribution is given by the eigenvector corresponding to the zeroth eigenvalue, 
\beq
p^{\infty}=\left(\frac{u+1}{2 s+4 u+2},\frac{s+u}{2 s+4 u+2},\frac{s+u}{2 s+4 u+2},\frac{u+1}{2 s+4 u+2}\right).
\eeq

These results allow us to calculate 
\begin{eqnarray}
\mu &= \frac{-e^{-2ut}(1-s)}{1+s -2u} + \frac{-e^{-(1+s)t}(\mu_0(1+s-2u) - (1-s))}{1+s -2u}\\
&=\mu_0e^{-(1+s)t} + \frac{1-s}{1+s-2u}\left(e^{-(1+s)t} - e^{-(1+s)t} \right).
\end{eqnarray}

\section{Model with feedback}\label{AppendixC}
The transition matrix for the model with feedback reads defined in Fig.~\ref{Fig1} B:
\begin{eqnarray}\label{LC}
\mathcal{L}=\left(
\begin{array}{cccc}
\alpha+s & -y  & -r  & 0\\
-\alpha  & y+r & 0   & -s\\
-s  & 0   & y+r & -\alpha\\
0   & -r  & -y  & s+\alpha\\
\end{array}
\right).
\end{eqnarray}

The detailed derivation of the steady state quantities and eigenvalues is given in Mancini et al~\cite{Mancini2015}. Here we just summarize the main results. The steady state probability distribution is:
\beq
\label{eq:steadystate_fb}
p^{\infty} = \frac{1}{2A} \{1+y, s +\alpha, s +\alpha, 1+y\},
\eeq
where we have defined $A$ and $\rho$ in Eqs.~\ref{eq:A} and ~\ref{eq:rho}.

The eigenvalues of the matrix in Eq,~\ref{LC} are $\{ \lambda_i \}=\{0,A,(A-\rho)/2,(A+\rho)/2\}$ and $\lambda=(A-\rho)/2$ is always the smallest eigenvalue. For a model with steady state initial conditions $\mu$ reads
\beqn
\label{eq:mu_fb}
&&\mu =
\exp\left(-\frac{A}{2 \lambda}\tau \right) \Big\{q \cosh\left(\frac{\rho}{2 \lambda}\tau\right)-\notag\\
&&\frac{\left[s^2-(1+y)^2-4\alpha+\alpha^2+2s(2y+\alpha)\right]}{A\rho}\sinh\left(\frac{\rho}{2 \lambda}\tau\right)\Big\}\notag,\\
\eeqn
with $q= {(1+y-s-\alpha)}/{A}$ and  the rescaled time $\tau = t \lambda$.

\begin{figure}
\begin{center}
\includegraphics[scale=0.9]{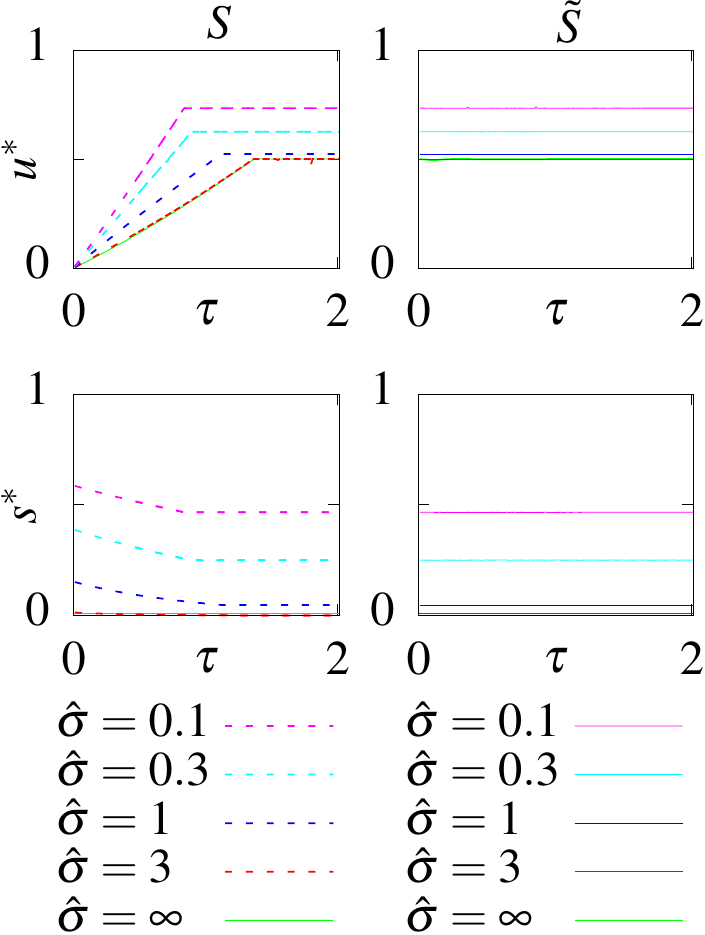}
\caption{\label{SIFIG_modelA}The optimal parameters as a function of the readout delay, $\tau$, for the models without feedback, $S$ and $\tilde{S}$, at different constrained steady state dissipation rates $\hat{\sigma}^{ss}$.}
\end{center}
\end{figure}

\begin{figure}
\begin{center}
\includegraphics[scale=0.9]{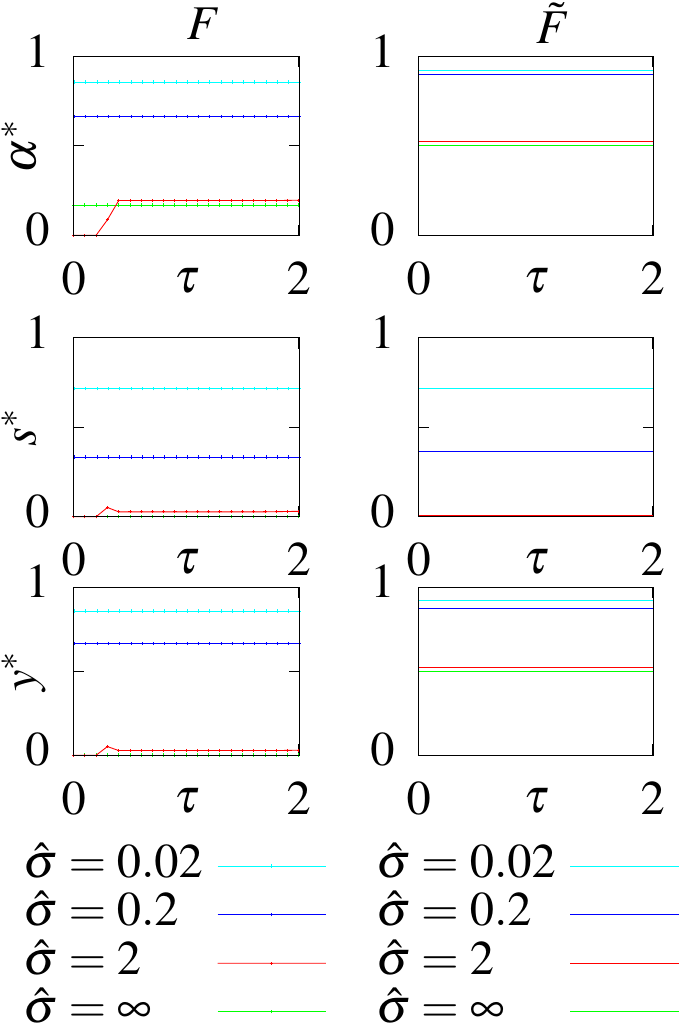}
\caption{\label{SIFIG_modelC}The optimal parameters as a function of the readout delay $\tau$ for models with feedback, $F$ and $\tilde{F}$, at different constrained steady state dissipation rates $\hat{\sigma}^{ss}$.}
\end{center}
\end{figure}

\section{Entropy Production Rate}\label{Appendix_entropyprid}
In this Appendix we present an alternative derivation of dissipation. 
We denote probability of state $i$  by $p_i$ and the \textit{entropy} of the distribution is defined as:
\begin{equation}\label{entropy}
S(t)=-\sum\limits_{i}^{}p_{i}(t)\log p_{i}(t).
\end{equation}

The entropy production rate formula is derived by differentiating the entropy with respect to time:
\begin{align*}
\dot{S}(t) &= -\sum\limits_{i}^{}\dot{p}_{i}(t)\log p_{i}(t) -\sum\limits_{i}^{}p_{i}(t)\frac{1}{p_{i}(t)}\dot{p}_{i}(t)\\
&= -\sum\limits_{i}^{}\dot{p}_{i}(t)\log p_{i}(t) -\left(\sum\limits_{i}^{}p_{i}(t)\right) ' .
\end{align*}
Denoting by $w_{ij}$ the transition rate from state $i$ to state $j$, we obtain $\dot{p}_i(t)=\sum\limits_{j\neq i}w_{ji}p_{j}(t)-w_{ij}p_{i}(t)$. We define $w_{ii}$ as $-\sum\limits_{j, j\neq i}w_{ij}$, so that we can write compactly $\dot{p}_i(t)=\sum\limits_{j}p_j(t)w_{ji}$ and the expression for $\dot{S}(t)$ becomes:
\begin{align}
\dot{S}(t)&= -\sum\limits_{i}^{} \left( \sum\limits_{j}^{} w_{ji}p_{j}(t) \right) \log p_{i}(t) - 0\nonumber\\
&= -\sum\limits_{i,j}w_{ji}p_{j}(t)\log p_{i}(t).\label{Sdot}
\end{align}
With the definition of $w_{ii}$, the terms $w_{ij}$ satisfy $\sum_{j}w_{ij}=0$. The following expression
$-\sum\limits_{i}^{}p_i(t)\log{p_i(t)}\sum\limits_{j}w_{ij}=-\sum\limits_{i,j}w_{ij}\log{p_i(t)}$ is then equal to zero and we subtract it form (\ref{Sdot}) to obtain a compact form:
\begin{eqnarray}
\dot{S}(t)&=& \left(\sum\limits_{i,j}p_{i}(t)w_{ij}\log{p_{i}(t)}-\sum\limits_{i,j}p_{i}(t)w_{ij}\log{p_{j}(t)}\right)= \\ \nonumber
 &&= \sum\limits_{i,j}p_{i}(t)w_{ij}\log \frac{p_{i}(t)}{p_{j}(t)}.
\end{eqnarray}

\noindent Further formula manipulation gives:
\begin{align}\label{epr}
\dot{S}(t) &= \frac{1}{2}\sum\limits_{i,j}p_{i}(t)w_{ij}\log \frac{p_{i}(t)}{p_{j}(t)} +\frac{1}{2}\sum\limits_{j,i}p_{j}(t)w_{ji}\log \frac{p_{j}(t)}{p_{i}(t)}\nonumber\\
&= \frac{1}{2} \sum\limits_{i,j}p_{i}(t)w_{ij}\log\frac{p_{i}(t)}{p_{j}(t)}-\frac{1}{2}\sum\limits_{j,i}p_{j}(t)w_{ji}\log \frac{p_{i}(t)}{p_{j}(t)}\nonumber\\
&= \frac{1}{2} \sum\limits_{i,j}\left(p_{i}(t)w_{ij}-p_{j}(t)w_{ji}\right)\log \frac{p_{i}(t)}{p_{j}(t)}\nonumber\\
&= \underbrace{\frac{1}{2} \sum\limits_{i,j}\left(p_{i}(t)w_{ij}-p_{j}(t)w_{ji}\right)\log \frac{w_{ji}}{w_{ij}}}_{\textit{entropy flow}} +\\ \nonumber
& \underbrace{\frac{1}{2} \sum\limits_{i,j}\left(p_{i}(t)w_{ij}-p_{j}(t)w_{ji}\right)\log \frac{p_{i}(t)w_{ij}}{p_{j}(t)w_{ji}}}_{\textit{entropy production rate}}.
\end{align}

The difference between the entropy production rate and the entropy flow, is the rate at which the whole entropy of a system changes. The entropy flow quantifies the flux of entropy from the system to the outside. In the steady state, as the entropy does not change, the two terms are equal, which means that the whole entropy produced by the system is dissipated.

The second underbracket of (Eq.~\ref{epr}) can be rewritten in the familiar form:
\begin{equation}
\sigma(t)= \sum\limits_{i,j}p_{i}(t)w_{ij}\log \frac{p_{i}(t)w_{ij}}{p_{j}(t)w_{ji}}.
\end{equation}

\section{Langevin description of bursty gene regulation}\label{Langevin}
A bursty model of transcription such as the one presented in section~\ref{Master_eq} can be written in a Langevin description introducing the frequency for the promoter to be in the activated state $n$~\cite{ Tkacik2011}:
\begin{eqnarray}
\frac{dn}{dt}&=& -ck_{_{+}}n -k_{_{-}} n + \xi_n\label{noise1},\\
\frac{dg}{dt}&=& Rn - \frac{1}{\tau}g + \xi_g, \label{noise2}
\end{eqnarray}
where the fluctuations are given by
\begin{eqnarray}
\langle\xi_n(t)\xi_n(t')\rangle&=& 2(k_+c(1-\bar{n})+k_-\bar{n})\delta(t-t'),\label{corr2}\\
\langle \xi_g(t) \xi_g(t')\rangle &=& 2(R\bar{n} + \bar{g}/\tau) \delta(t-t').\label{corr1}
\end{eqnarray}
These equations describe the fluctuations of the promoter state and the protein concentration $g$ around the equilibrium solution
$\left(\overline{n},\overline{g}\right)=\left(\frac{k_{_{+}}
  c}{k_{_{-}}+k_{_{+}} c},\frac{k_{_{+}} c R \tau}{k_{_{-}}+k_{_{+}}
  c}\right)$. In order to lighten notations, we have used $(n,g)$ instead of
the standard form $(\delta n, \delta g)$ to describe fluctuations. Eqs.~(\ref{noise1}-\ref{noise2}) can be recast into the matrix form form ${\bf \dot{X}}= -A {\bf X}+ \boldsymbol\xi$ with

\begin{equation}\label{AAnoise}
A=\left(\begin{array}{cc}
ck_{_{+}}+k_{_{-}} & 0\\
-R & \frac{1}{\tau}
\end{array}\right),
\end{equation}
and the noise correlation matrix \\$\langle\boldsymbol\xi(t) \boldsymbol\xi(t')\rangle = 2D\delta(t-t')$ is 
\begin{equation}\label{DDnoise}
D=\left(\begin{array}{cc}
k_+c(1-\bar{n})+k_-\bar{n}& 0\\
0 & (R\bar{n} + \bar{g}/\tau)
\end{array}\right).
\end{equation}
The correlation matrix $\Sigma$ can be computed with standard methods~\cite{Tkacik2012}, by  inverting the relation $D=A\Sigma+\Sigma A^{t}$:
\begin{eqnarray}\label{corr_ng}
\Sigma&=&\left(\begin{array}{cc}\langle n n \rangle & \langle n g \rangle\\
\langle g n \rangle & \langle g g \rangle\end{array}\right)=
\frac{1}{(c k_{_{+}}+k_{_{-}})^2} \cdot \\ \nonumber
&&\left(
\begin{array}{cc}
 2 c k_{_{-}} k_{_{+}} & \frac{2 c
   k_{_{-}} k_{_{+}} R \tau }{( (c k_{_{+}}
   \tau +k_{_{-}} \tau +1)} \\
 \frac{2 c k_{_{-}} k_{_{+}} R \tau }{ (c
   k_{_{+}} \tau +k_{_{-}} \tau +1)} & \frac{2 c k_{_{+}} R \tau 
   \left(\tau  \left((c k_{_{+}}+k_{_{-}})^2+k_{_{-}} R\right)+c
   k_{_{+}}+k_{_{-}}\right)}{ (c k_{_{+}}
   \tau +k_{_{-}} \tau +1)} \\
\end{array}
\right).
\end{eqnarray}
  
\subsection{Entropy production}\label{sec:entr_prod}

The probability of a trajectory of a multivariate Langevin process can
be calculated via the Onsager-Machlup formalism. Using this
probability as starting point, the dissipation can be exactly derived
(see Ref.~\cite{Puglisi_2009}, where the computation is done in
detail and in a self-contained fashion).  For the case of symmetric
variables under time reversal the entropy production can be written in a
compact form, where we have the index $k$ run over all the variables:
\begin{equation} 
W(t) = \sum_{k}D^{-1}_{kk}\int_{0}^{t}ds \left(A {\bf X}\right)_{k} \dot{X}_{k} \label{formulone2},
\end{equation}
In our case, by using Eqs.~(\ref{AAnoise}) and (\ref{DDnoise}), one has 
\begin{eqnarray}
W(t)&=&D^{-1}_{nn}\int_{0}^{t}dt'\,(ck_{_{+}}n(t') -k_{_{-}} n(t'))\dot{n}(t')+ \nonumber\\
&&D^{-1}_{gg}\int_{0}^{t} dt'\,(Rn(t') - \frac{1}{\tau}g(t'))\dot{g}(t').\label{entr_prod_model2}
\end{eqnarray}

\begin{figure}
\begin{center}
\includegraphics[scale=1.3]{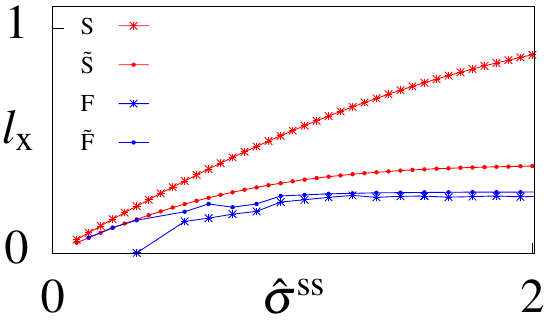}
\caption{\label{SIFIG_learningrate}The learning rate for the output variable $x$ as a function of the rescaled steady state dissipation, $\hat{\sigma}^{\text{ss}}$, calculated at steady state for models with ($F$ and $\tilde{F}$) and without feedback ($S$ and $\tilde{S}$). Models $\tilde{S}$ and $\tilde{F}$ have optimized initial conditions (that do not enter this calculations except for the optimal parameters) and models $S$ and $F$ are constrained to have initial conditions in steady state. }
\end{center}
\end{figure}

Eq.~\eqref{entr_prod_model2} can be simplified by considering that all  terms which are exact derivatives are not extensive in time (terms like $\int_{0}^{t} dt' n(t') \dot{n}(t') =\frac{1}{2}\left(n^{2}(t)-n^{2}(0)\right)$ or its equivalent in $g$ can be neglected in the large $t$ limit. Moreover all the steady state correlations are time translational invariant, i.e.  $\left<\int_{0}^{t} dt' n(t') \dot{g}(t')\right>\equiv t \left<n \dot{g}\right> $. As a consequence, the dissipation becomes:
\begin{equation}
\sigma^{LE}=\lim_{t\to \infty}\frac{\left<W_{t}\right>}{t}=\frac{R}{R\overline{n}+\frac{\overline{g}}{\tau}}\left<n\dot{g}\right>. \label{entr}
\end{equation}

The correlation $\left<n\dot{g}\right>$ in Eq.~\eqref{entr} can be
computed by replacing $\dot{g}$ with Eq.~\eqref{noise2},   yielding
$\left<n\dot{g}\right>=R
\left<nn\right>-\frac{1}{\tau}\left<ng\right>$. Substituting
this expression into Eq.~\eqref{corr_ng} we obtain:
\begin{equation}\label{dissipation2}
\sigma^{LE}(c)=\frac{R\tau (1-ck_{+}\tau_{s})}{\tau+\tau_s},
\end{equation}
where $\tau_s= ({c  k_{_{+}}+k_{_{-}}})^{-1}$.
Note that in
the limit $\tau_{s}\to 0$ the dissipation is not dependent on $c$ and
equal to $\sigma^{LE}_0={R}/{(1+cK)}$, where $K$ is equal to
$k_{_{+}}/k_{_{-}}$. Moreover, for $K\to \infty$ (which corresponds to no flux
to the inactive state, $k_-\to 0$) the dissipation vanishes like in
the master equation formulation (Eq.~(\ref{fast_diss})).\\

As final remark, we note that a Langevin formulation is a coarse
grained description of the Master equation approach described in
Sec.~\ref{Master_eq}. This kind of coarse graining procedure integrates away
degrees of freedom which can carry non-equilibrium currents and can
lead to lower values of dissipation~\cite{Crisanti2012,
  Puglisi2015, Busiello2019}. For instance, consider the limit $R_0
\approx \epsilon$ small but finite, Eq.~\eqref{fast_diss} becomes
$\sigma^{ME}=cK(1+R_1)/(1+cK)^2 \log \frac{R_1}{\epsilon}$ and one
finds $\sigma^{ME}>\sigma^{LE}$. 
                                                                                                                                      
\section{Learning Rate}~\label{learning_rate}
Lastly, following Barato et al~\cite{Barrato2015} we consider the learning rate in steady state (we limit ourselves to the steady state discussion since it allows us to get analytical intuition)
\begin{equation}
l_{x}=-\sum_{i} {p^{\infty}_{i}}\sum_{i\neq j}w_{ij}\log\frac{p^{\infty}_{i}}{p_{j}^{\infty}},
\end{equation}
which was defined to describe the rate at which the output $x$ learns about the dynamics of the stochastic input $z$. For our system, the learning rate is explicitly given by
\begin{eqnarray}\label{learning_spec}
-l_{x}&=&p^{\infty}_{1}w_{13}\log\frac{p^{\infty}_{1}}{p^{\infty}_{3}}+ p^{\infty}_{3}w_{31}\log\frac{p^{\infty}_{3}}{p^{\infty}_{1}}\\ \nonumber
&&p^{\infty}_{2}w_{24}\log\frac{p^{\infty}_{2}}{p^{\infty}_{4}} +
p^{\infty}_{4}w_{42}\log\frac{p^{\infty}_{4}}{p^{\infty}_{2}},\\
\end{eqnarray}
and is bounded by $\sigma_x$ defined as:
\begin{eqnarray}
\sigma_{x}&=&(p^{\infty}_{1}w_{13}-p^{\infty}_{3}w_{31})w_{13}\log\frac{w_{13}}{w_{31}}\\ \nonumber
&&(p^{\infty}_{2}w_{24}-p^{\infty}_{4}w_{42})\log\frac{w_{24}}{w_{42}}.
\end{eqnarray}

For the models without feedback ($S$ and $\tilde{S}$) the learning rate is:
\begin{equation}
l_{x}=\frac{u(s-1)}{s+2u+1}\log\frac{u+s}{u+1}.\\
\end{equation}
In models without feedback $w_{12}=w_{21}$ and $w_{34}=w_{43}$ and the steady state dissipation rate comes only from the output, $x$ ($\sigma_z=0$ and $\sigma_x=\sigma^\text{ss}$) and is given by Eq.~\ref{ss_ent_modelA}, such that
\begin{equation}
\eta=\frac{l_{x}}{\sigma_x}={\log\frac{u+s}{u+1}}/{\log{s}} \leq 1.\\
\end{equation}

For models with feedback ($F$ and $\tilde{F}$) the learning rate is harder to interpret since the input no longer changes independently of the output. Formally we can still calculate the quantity in Eq.~\ref{learning_spec} as
\begin{equation}
l_{x}=\frac{y(s-\alpha)}{\alpha+s+y+1}\log \frac{y+1}{\alpha+1},\\
\end{equation}
and 
\begin{equation}
\sigma_x=\frac{y(s-\alpha)}{\alpha+s+y+1}\log s.\\
\end{equation}
The informational efficiency is:
\begin{equation}
\eta=\frac{y+1}{\alpha+1}/\log{s},\\
\end{equation}
which is bounded by $1$ only if $s y \leq \alpha$ (see Fig.~\ref{SIFIG_learningrate}).


\begin{thebibliography}{77}%
\makeatletter
\providecommand \@ifxundefined [1]{%
 \@ifx{#1\undefined}
}%
\providecommand \@ifnum [1]{%
 \ifnum #1\expandafter \@firstoftwo
 \else \expandafter \@secondoftwo
 \fi
}%
\providecommand \@ifx [1]{%
 \ifx #1\expandafter \@firstoftwo
 \else \expandafter \@secondoftwo
 \fi
}%
\providecommand \natexlab [1]{#1}%
\providecommand \enquote  [1]{``#1''}%
\providecommand \bibnamefont  [1]{#1}%
\providecommand \bibfnamefont [1]{#1}%
\providecommand \citenamefont [1]{#1}%
\providecommand \href@noop [0]{\@secondoftwo}%
\providecommand \href [0]{\begingroup \@sanitize@url \@href}%
\providecommand \@href[1]{\@@startlink{#1}\@@href}%
\providecommand \@@href[1]{\endgroup#1\@@endlink}%
\providecommand \@sanitize@url [0]{\catcode `\\12\catcode `\$12\catcode
  `\&12\catcode `\#12\catcode `\^12\catcode `\_12\catcode `\%12\relax}%
\providecommand \@@startlink[1]{}%
\providecommand \@@endlink[0]{}%
\providecommand \url  [0]{\begingroup\@sanitize@url \@url }%
\providecommand \@url [1]{\endgroup\@href {#1}{\urlprefix }}%
\providecommand \urlprefix  [0]{URL }%
\providecommand \Eprint [0]{\href }%
\providecommand \doibase [0]{http://dx.doi.org/}%
\providecommand \selectlanguage [0]{\@gobble}%
\providecommand \bibinfo  [0]{\@secondoftwo}%
\providecommand \bibfield  [0]{\@secondoftwo}%
\providecommand \translation [1]{[#1]}%
\providecommand \BibitemOpen [0]{}%
\providecommand \bibitemStop [0]{}%
\providecommand \bibitemNoStop [0]{.\EOS\space}%
\providecommand \EOS [0]{\spacefactor3000\relax}%
\providecommand \BibitemShut  [1]{\csname bibitem#1\endcsname}%
\let\auto@bib@innerbib\@empty
\bibitem [{\citenamefont {Bialek}(2012)}]{Bialek_book}%
  \BibitemOpen
  \bibfield  {author} {\bibinfo {author} {\bibfnamefont {W.}~\bibnamefont
  {Bialek}},\ }\href@noop {} {\emph {\bibinfo {title} {{Biophysics}}}}\
  (\bibinfo  {publisher} {Princeton University Press},\ \bibinfo {address}
  {Princeton},\ \bibinfo {year} {2012})\BibitemShut {NoStop}%
\bibitem [{\citenamefont {Alon}(2006)}]{Alon2006}%
  \BibitemOpen
  \bibfield  {author} {\bibinfo {author} {\bibfnamefont {U.}~\bibnamefont
  {Alon}},\ }\href@noop {} {\emph {\bibinfo {title} {An Introduction to Systems
  Biology: Design Principles of Biological Circuits}}}\ (\bibinfo  {publisher}
  {Chapman \& Hall},\ \bibinfo {year} {2006})\BibitemShut {NoStop}%
\bibitem [{\citenamefont {Phillips}\ \emph {et~al.}()\citenamefont {Phillips},
  \citenamefont {Kondev}, \citenamefont {Theriot},\ and\ \citenamefont
  {Garcia}}]{Phillips_book}%
  \BibitemOpen
  \bibfield  {author} {\bibinfo {author} {\bibfnamefont {R.}~\bibnamefont
  {Phillips}}, \bibinfo {author} {\bibfnamefont {J.}~\bibnamefont {Kondev}},
  \bibinfo {author} {\bibfnamefont {J.}~\bibnamefont {Theriot}}, \ and\
  \bibinfo {author} {\bibfnamefont {H.}~\bibnamefont {Garcia}},\ }\href@noop {}
  {\emph {\bibinfo {title} {{Physical Biology of the Cell}}}}\ (\bibinfo
  {publisher} {Garland Science})\ p.\ \bibinfo {pages} {2012}\BibitemShut
  {NoStop}%
\bibitem [{\citenamefont {Lan}\ \emph {et~al.}(2012)\citenamefont {Lan},
  \citenamefont {Sartori}, \citenamefont {Neumann}, \citenamefont {Sourjik},\
  and\ \citenamefont {Tu}}]{Lan2012}%
  \BibitemOpen
  \bibfield  {author} {\bibinfo {author} {\bibfnamefont {G.}~\bibnamefont
  {Lan}}, \bibinfo {author} {\bibfnamefont {P.}~\bibnamefont {Sartori}},
  \bibinfo {author} {\bibfnamefont {S.}~\bibnamefont {Neumann}}, \bibinfo
  {author} {\bibfnamefont {V.}~\bibnamefont {Sourjik}}, \ and\ \bibinfo
  {author} {\bibfnamefont {Y.}~\bibnamefont {Tu}},\ }\href@noop {} {\bibfield
  {journal} {\bibinfo  {journal} {Nature Physics}\ }\textbf {\bibinfo {volume}
  {8}},\ \bibinfo {pages} {422} (\bibinfo {year} {2012})}\BibitemShut {NoStop}%
\bibitem [{\citenamefont {Mehta}\ and\ \citenamefont
  {Schwab}(2012)}]{Mehta2012}%
  \BibitemOpen
  \bibfield  {author} {\bibinfo {author} {\bibfnamefont {P.}~\bibnamefont
  {Mehta}}\ and\ \bibinfo {author} {\bibfnamefont {D.~J.}\ \bibnamefont
  {Schwab}},\ }\href@noop {} {\bibfield  {journal} {\bibinfo  {journal}
  {Proceedings of the National Academy of Sciences of the United States of
  America}\ }\textbf {\bibinfo {volume} {109}},\ \bibinfo {pages} {17978–17982}
  (\bibinfo {year} {2012})}\BibitemShut {NoStop}%
\bibitem [{\citenamefont {Cao}\ \emph {et~al.}(2015)\citenamefont {Cao},
  \citenamefont {Wang}, \citenamefont {Ouyang},\ and\ \citenamefont
  {Tu}}]{Cao2015}%
  \BibitemOpen
  \bibfield  {author} {\bibinfo {author} {\bibfnamefont {Y.}~\bibnamefont
  {Cao}}, \bibinfo {author} {\bibfnamefont {H.}~\bibnamefont {Wang}}, \bibinfo
  {author} {\bibfnamefont {Q.}~\bibnamefont {Ouyang}}, \ and\ \bibinfo {author}
  {\bibfnamefont {Y.}~\bibnamefont {Tu}},\ }\href {\doibase 10.1038/NPHYS3412}
  {\bibfield  {journal} {\bibinfo  {journal} {Nature Physics}\ ,\ \bibinfo
  {pages} {1}} (\bibinfo {year} {2015})}\BibitemShut {NoStop}%
\bibitem [{\citenamefont {Milo}\ and\ \citenamefont
  {Phillips}(2015)}]{MiloPhillips}%
  \BibitemOpen
  \bibfield  {author} {\bibinfo {author} {\bibfnamefont {R.}~\bibnamefont
  {Milo}}\ and\ \bibinfo {author} {\bibfnamefont {R.}~\bibnamefont
  {Phillips}},\ }\href@noop {} {\emph {\bibinfo {title} {{Cell Biology by the
  Numbers}}}}\ (\bibinfo  {publisher} {Garland Science},\ \bibinfo {year}
  {2015})\ p.\ \bibinfo {pages} {2015}\BibitemShut {NoStop}%
\bibitem [{\citenamefont {Moran}\ \emph {et~al.}(2010)\citenamefont {Moran},
  \citenamefont {Phillips},\ and\ \citenamefont {Milo}}]{Moran2010}%
  \BibitemOpen
  \bibfield  {author} {\bibinfo {author} {\bibfnamefont {U.}~\bibnamefont
  {Moran}}, \bibinfo {author} {\bibfnamefont {R.}~\bibnamefont {Phillips}}, \
  and\ \bibinfo {author} {\bibfnamefont {R.}~\bibnamefont {Milo}},\ }\href
  {\doibase 10.1016/j.cell.2010.06.019} {\bibfield  {journal} {\bibinfo
  {journal} {Cell}\ }\textbf {\bibinfo {volume} {141}},\ \bibinfo {pages}
  {1262} (\bibinfo {year} {2010})}\BibitemShut {NoStop}%
\bibitem [{\citenamefont {Seifert}(2012)}]{Seifert2012}%
  \BibitemOpen
  \bibfield  {author} {\bibinfo {author} {\bibfnamefont {U.}~\bibnamefont
  {Seifert}},\ }\href@noop {} {\bibfield  {journal} {\bibinfo  {journal}
  {Reports on Progress in Physics}\ }\textbf {\bibinfo {volume} {75}},\
  \bibinfo {pages} {126001} (\bibinfo {year} {2012})}\BibitemShut {NoStop}%
\bibitem [{\citenamefont {Still}\ \emph {et~al.}(2012)\citenamefont {Still},
  \citenamefont {Sivak}, \citenamefont {Bell},\ and\ \citenamefont
  {Crooks}}]{Still2012}%
  \BibitemOpen
  \bibfield  {author} {\bibinfo {author} {\bibfnamefont {S.}~\bibnamefont
  {Still}}, \bibinfo {author} {\bibfnamefont {D.~A.}\ \bibnamefont {Sivak}},
  \bibinfo {author} {\bibfnamefont {A.~J.}\ \bibnamefont {Bell}}, \ and\
  \bibinfo {author} {\bibfnamefont {G.~E.}\ \bibnamefont {Crooks}},\
  }\href@noop {} {\bibfield  {journal} {\bibinfo  {journal} {Physical Review
  Letters}\ }\textbf {\bibinfo {volume} {109}},\ \bibinfo {pages} {120604}
  (\bibinfo {year} {2012})}\BibitemShut {NoStop}%
\bibitem [{\citenamefont {Ouldridge}\ \emph {et~al.}(2017)\citenamefont
  {Ouldridge}, \citenamefont {Govern},\ and\ \citenamefont
  {Rein}}]{Ouldridge2017}%
  \BibitemOpen
  \bibfield  {author} {\bibinfo {author} {\bibfnamefont {T.~E.}\ \bibnamefont
  {Ouldridge}}, \bibinfo {author} {\bibfnamefont {C.~C.}\ \bibnamefont
  {Govern}}, \ and\ \bibinfo {author} {\bibfnamefont {P.}~\bibnamefont
  {Rein}},\ }\href {\doibase 10.1103/PhysRevX.7.021004} {\bibfield  {journal}
  {\bibinfo  {journal} {Physical Review X}\ }\textbf {\bibinfo {volume}
  {021004}},\ \bibinfo {pages} {1} (\bibinfo {year} {2017})}\BibitemShut
  {NoStop}%
\bibitem [{\citenamefont {Rein}\ \emph {et~al.}(2016)\citenamefont {Rein},
  \citenamefont {Becker}, \citenamefont {Ouldridge},\ and\ \citenamefont
  {Mugler}}]{tenwolde2016}%
  \BibitemOpen
  \bibfield  {author} {\bibinfo {author} {\bibfnamefont {P.}~\bibnamefont
  {Rein}}, \bibinfo {author} {\bibfnamefont {N.~B.}\ \bibnamefont {Becker}},
  \bibinfo {author} {\bibfnamefont {T.~E.}\ \bibnamefont {Ouldridge}}, \ and\
  \bibinfo {author} {\bibfnamefont {A.}~\bibnamefont {Mugler}},\ }\href
  {\doibase 10.1007/s10955-015-1440-5} {\bibfield  {journal} {\bibinfo
  {journal} {Journal of Statistical Physics}\ }\textbf {\bibinfo {volume}
  {162}},\ \bibinfo {pages} {1395} (\bibinfo {year} {2016})}\BibitemShut
  {NoStop}%
\bibitem [{\citenamefont {Sagawa}\ and\ \citenamefont
  {Ito}(2015)}]{itosagawa2015}%
  \BibitemOpen
  \bibfield  {author} {\bibinfo {author} {\bibfnamefont {T.}~\bibnamefont
  {Sagawa}}\ and\ \bibinfo {author} {\bibfnamefont {S.}~\bibnamefont {Ito}},\
  }\href {\doibase 10.1038/ncomms8498} {\bibfield  {journal} {\bibinfo
  {journal} {Nature Communications}\ ,\ \bibinfo {pages} {2}} (\bibinfo {year}
  {2015})}\BibitemShut {NoStop}%
\bibitem [{\citenamefont {Hopfield}(1974)}]{Hopfield1974}%
  \BibitemOpen
  \bibfield  {author} {\bibinfo {author} {\bibfnamefont {J.}~\bibnamefont
  {Hopfield}},\ }\href@noop {} {\bibfield  {journal} {\bibinfo  {journal}
  {Proceedings of the National Academy of Sciences of the United States of
  America}\ }\textbf {\bibinfo {volume} {71}},\ \bibinfo {pages} {4135}
  (\bibinfo {year} {1974})}\BibitemShut {NoStop}%
\bibitem [{\citenamefont {Ninio}(1975)}]{Ninio1975}%
  \BibitemOpen
  \bibfield  {author} {\bibinfo {author} {\bibfnamefont {J.}~\bibnamefont
  {Ninio}},\ }\href@noop {} {\bibfield  {journal} {\bibinfo  {journal}
  {Biochimie}\ }\textbf {\bibinfo {volume} {57}},\ \bibinfo {pages} {587}
  (\bibinfo {year} {1975})}\BibitemShut {NoStop}%
\bibitem [{\citenamefont {Barato}\ \emph {et~al.}(2013)\citenamefont {Barato},
  \citenamefont {Hartich},\ and\ \citenamefont {Seifert}}]{Barato2013}%
  \BibitemOpen
  \bibfield  {author} {\bibinfo {author} {\bibfnamefont {A.~C.}\ \bibnamefont
  {Barato}}, \bibinfo {author} {\bibfnamefont {D.}~\bibnamefont {Hartich}}, \
  and\ \bibinfo {author} {\bibfnamefont {U.}~\bibnamefont {Seifert}},\
  }\href@noop {} {\bibfield  {journal} {\bibinfo  {journal} {Physical Review
  E}\ }\textbf {\bibinfo {volume} {87}},\ \bibinfo {pages} {042104} (\bibinfo
  {year} {2013})}\BibitemShut {NoStop}%
\bibitem [{\citenamefont {Barato}\ \emph {et~al.}(2014)\citenamefont {Barato},
  \citenamefont {Hartich},\ and\ \citenamefont {Seifert}}]{Barato2014}%
  \BibitemOpen
  \bibfield  {author} {\bibinfo {author} {\bibfnamefont {A.~C.}\ \bibnamefont
  {Barato}}, \bibinfo {author} {\bibfnamefont {D.}~\bibnamefont {Hartich}}, \
  and\ \bibinfo {author} {\bibfnamefont {U.}~\bibnamefont {Seifert}},\
  }\href@noop {} {\bibfield  {journal} {\bibinfo  {journal} {New Journal of
  Physics}\ }\textbf {\bibinfo {volume} {16}},\ \bibinfo {pages} {103024}
  (\bibinfo {year} {2014})}\BibitemShut {NoStop}%
\bibitem [{\citenamefont {Bo}\ \emph {et~al.}(2015)\citenamefont {Bo},
  \citenamefont {Giudice},\ and\ \citenamefont {Celani}}]{Bo2015}%
  \BibitemOpen
  \bibfield  {author} {\bibinfo {author} {\bibfnamefont {S.}~\bibnamefont
  {Bo}}, \bibinfo {author} {\bibfnamefont {M.~D.}\ \bibnamefont {Giudice}}, \
  and\ \bibinfo {author} {\bibfnamefont {A.}~\bibnamefont {Celani}},\
  }\href@noop {} {\bibfield  {journal} {\bibinfo  {journal} {Journal of
  Statistical Mechanics: Theory and Experiment}\ }\textbf {\bibinfo {volume}
  {2015}},\ \bibinfo {pages} {P01014} (\bibinfo {year} {2015})}\BibitemShut
  {NoStop}%
\bibitem [{\citenamefont {Govern}\ and\ \citenamefont {ten
  Wolde}(2014)}]{Govern2014}%
  \BibitemOpen
  \bibfield  {author} {\bibinfo {author} {\bibfnamefont {C.~C.}\ \bibnamefont
  {Govern}}\ and\ \bibinfo {author} {\bibfnamefont {P.~R.}\ \bibnamefont {ten
  Wolde}},\ }\href@noop {} {\bibfield  {journal} {\bibinfo  {journal} {Physical
  Review Letters}\ }\textbf {\bibinfo {volume} {113}},\ \bibinfo {pages}
  {258102} (\bibinfo {year} {2014})}\BibitemShut {NoStop}%
\bibitem [{\citenamefont {Barato}\ and\ \citenamefont
  {Seifert}(2015)}]{Barrato2015}%
  \BibitemOpen
  \bibfield  {author} {\bibinfo {author} {\bibfnamefont {A.~C.}\ \bibnamefont
  {Barato}}\ and\ \bibinfo {author} {\bibfnamefont {U.}~\bibnamefont
  {Seifert}},\ }\href {\doibase 10.1103/PhysRevLett.} {\bibfield  {journal}
  {\bibinfo  {journal} {Physical Review Letters}\ }\textbf {\bibinfo {volume}
  {158101}},\ \bibinfo {pages} {1} (\bibinfo {year} {2015})}\BibitemShut
  {NoStop}%
\bibitem [{\citenamefont {Brittain}\ \emph {et~al.}(2017)\citenamefont
  {Brittain}, \citenamefont {Jones},\ and\ \citenamefont
  {Ouldridge}}]{Brittain2017}%
  \BibitemOpen
  \bibfield  {author} {\bibinfo {author} {\bibfnamefont {R.~A.}\ \bibnamefont
  {Brittain}}, \bibinfo {author} {\bibfnamefont {N.~S.}\ \bibnamefont {Jones}},
  \ and\ \bibinfo {author} {\bibfnamefont {T.~E.}\ \bibnamefont {Ouldridge}},\
  }\href@noop {} {\bibfield  {journal} {\bibinfo  {journal} {Journal of
  Statistical Mechanics: Theory and Experiment}\ } (\bibinfo {year}
  {2017})}\BibitemShut {NoStop}%
\bibitem [{\citenamefont {Goldt}\ and\ \citenamefont
  {Seifert}(2017)}]{Goldt2017}%
  \BibitemOpen
  \bibfield  {author} {\bibinfo {author} {\bibfnamefont {S.}~\bibnamefont
  {Goldt}}\ and\ \bibinfo {author} {\bibfnamefont {U.}~\bibnamefont
  {Seifert}},\ }\href {\doibase 10.1103/PhysRevLett.118.010601} {\bibfield
  {journal} {\bibinfo  {journal} {Physical Review Letters}\ }\textbf {\bibinfo
  {volume} {010601}},\ \bibinfo {pages} {1} (\bibinfo {year}
  {2017})}\BibitemShut {NoStop}%
\bibitem [{\citenamefont {Parrondo}\ \emph {et~al.}(2015)\citenamefont
  {Parrondo}, \citenamefont {Horowitz},\ and\ \citenamefont
  {Sagawa}}]{Parrando2015}%
  \BibitemOpen
  \bibfield  {author} {\bibinfo {author} {\bibfnamefont {J.~M.~R.}\
  \bibnamefont {Parrondo}}, \bibinfo {author} {\bibfnamefont {J.~M.}\
  \bibnamefont {Horowitz}}, \ and\ \bibinfo {author} {\bibfnamefont
  {T.}~\bibnamefont {Sagawa}},\ }\href {\doibase 10.1038/NPHYS3230} {\bibfield
  {journal} {\bibinfo  {journal} {Nature Physics}\ }\textbf {\bibinfo {volume}
  {11}},\ \bibinfo {pages} {131} (\bibinfo {year} {2015})}\BibitemShut
  {NoStop}%
\bibitem [{\citenamefont {Becker}\ \emph {et~al.}(2013)\citenamefont {Becker},
  \citenamefont {Mugler},\ and\ \citenamefont {ten Wolde}}]{Becker2013}%
  \BibitemOpen
  \bibfield  {author} {\bibinfo {author} {\bibfnamefont {N.~B.}\ \bibnamefont
  {Becker}}, \bibinfo {author} {\bibfnamefont {A.}~\bibnamefont {Mugler}}, \
  and\ \bibinfo {author} {\bibfnamefont {P.~R.}\ \bibnamefont {ten Wolde}},\
  }\href@noop {} {\enquote {\bibinfo {title} {Prediction and dissipation in
  biochemical sensing},}\ } (\bibinfo {year} {2013}),\ \bibinfo {note}
  {http://arxiv.org/abs/1312.5625}\BibitemShut {NoStop}%
\bibitem [{\citenamefont {Horowitz}\ and\ \citenamefont
  {Esposito}(2014)}]{Horowitz2014}%
  \BibitemOpen
  \bibfield  {author} {\bibinfo {author} {\bibfnamefont {J.~M.}\ \bibnamefont
  {Horowitz}}\ and\ \bibinfo {author} {\bibfnamefont {M.}~\bibnamefont
  {Esposito}},\ }\href {\doibase 10.1103/PhysRevX.4.031015} {\bibfield
  {journal} {\bibinfo  {journal} {Physical Review X}\ }\textbf {\bibinfo
  {volume} {4}},\ \bibinfo {pages} {031015} (\bibinfo {year}
  {2014})}\BibitemShut {NoStop}%
\bibitem [{\citenamefont {Allahverdyan}\ \emph {et~al.}(2009)\citenamefont
  {Allahverdyan}, \citenamefont {Janzing},\ and\ \citenamefont
  {Mahler}}]{Allahverdyan2009}%
  \BibitemOpen
  \bibfield  {author} {\bibinfo {author} {\bibfnamefont {A.~E.}\ \bibnamefont
  {Allahverdyan}}, \bibinfo {author} {\bibfnamefont {D.}~\bibnamefont
  {Janzing}}, \ and\ \bibinfo {author} {\bibfnamefont {G.}~\bibnamefont
  {Mahler}},\ }\href {\doibase 10.1088/1742-5468/2009/09/P09011} {\bibfield
  {journal} {\bibinfo  {journal} {Journal of Statistical Mechanics: Theory and
  Experiment}\ }\textbf {\bibinfo {volume} {2009}},\ \bibinfo {pages} {P09011}
  (\bibinfo {year} {2009})}\BibitemShut {NoStop}%
\bibitem [{\citenamefont {Sartori}\ \emph {et~al.}(2014)\citenamefont
  {Sartori}, \citenamefont {Granger}, \citenamefont {Lee},\ and\ \citenamefont
  {Horowitz}}]{Sartori2014}%
  \BibitemOpen
  \bibfield  {author} {\bibinfo {author} {\bibfnamefont {P.}~\bibnamefont
  {Sartori}}, \bibinfo {author} {\bibfnamefont {L.}~\bibnamefont {Granger}},
  \bibinfo {author} {\bibfnamefont {C.~F.}\ \bibnamefont {Lee}}, \ and\
  \bibinfo {author} {\bibfnamefont {J.~M.}\ \bibnamefont {Horowitz}},\
  }\href@noop {} {\bibfield  {journal} {\bibinfo  {journal} {PLOS Computational
  Biology}\ }\textbf {\bibinfo {volume} {10}},\ \bibinfo {pages} {e1003974}
  (\bibinfo {year} {2014})}\BibitemShut {NoStop}%
\bibitem [{\citenamefont {Hartich}\ \emph {et~al.}(2016)\citenamefont
  {Hartich}, \citenamefont {Barato},\ and\ \citenamefont
  {Seifert}}]{Hartich2016}%
  \BibitemOpen
  \bibfield  {author} {\bibinfo {author} {\bibfnamefont {D.}~\bibnamefont
  {Hartich}}, \bibinfo {author} {\bibfnamefont {A.~C.}\ \bibnamefont {Barato}},
  \ and\ \bibinfo {author} {\bibfnamefont {U.}~\bibnamefont {Seifert}},\ }\href
  {\doibase 10.1103/PhysRevE.93.022116} {\bibfield  {journal} {\bibinfo
  {journal} {Physical Review E}\ }\textbf {\bibinfo {volume} {022116}},\
  \bibinfo {pages} {1} (\bibinfo {year} {2016})}\BibitemShut {NoStop}%
\bibitem [{\citenamefont {Falasco}\ \emph {et~al.}(2018)\citenamefont
  {Falasco}, \citenamefont {Rao},\ and\ \citenamefont
  {Esposito}}]{Falasco2018}%
  \BibitemOpen
  \bibfield  {author} {\bibinfo {author} {\bibfnamefont {G.}~\bibnamefont
  {Falasco}}, \bibinfo {author} {\bibfnamefont {R.}~\bibnamefont {Rao}}, \ and\
  \bibinfo {author} {\bibfnamefont {M.}~\bibnamefont {Esposito}},\ }\href
  {\doibase 10.1103/PhysRevLett.121.108301} {\bibfield  {journal} {\bibinfo
  {journal} {Physical Review Letters}\ }\textbf {\bibinfo {volume} {121}},\
  \bibinfo {pages} {108301} (\bibinfo {year} {2018})}\BibitemShut {NoStop}%
\bibitem [{\citenamefont {Tka\v{c}ik}\ \emph {et~al.}(2008)\citenamefont
  {Tka\v{c}ik}, \citenamefont {Callan},\ and\ \citenamefont
  {Bialek}}]{Tkavcik2008}%
  \BibitemOpen
  \bibfield  {author} {\bibinfo {author} {\bibfnamefont {G.}~\bibnamefont
  {Tka\v{c}ik}}, \bibinfo {author} {\bibfnamefont {C.~G.}\ \bibnamefont
  {Callan}}, \ and\ \bibinfo {author} {\bibfnamefont {W.}~\bibnamefont
  {Bialek}},\ }\href@noop {} {\bibfield  {journal} {\bibinfo  {journal}
  {Proceedings of the National Academy of Sciences of the United States of
  America}\ }\textbf {\bibinfo {volume} {105}},\ \bibinfo {pages} {12265}
  (\bibinfo {year} {2008})}\BibitemShut {NoStop}%
\bibitem [{\citenamefont {Tka\v{c}ik}\ and\ \citenamefont
  {Walczak}(2011)}]{Tkacik2011}%
  \BibitemOpen
  \bibfield  {author} {\bibinfo {author} {\bibfnamefont {G.}~\bibnamefont
  {Tka\v{c}ik}}\ and\ \bibinfo {author} {\bibfnamefont {A.~M.}\ \bibnamefont
  {Walczak}},\ }\href {\doibase 10.1088/0953-8984/23/15/153102} {\bibfield
  {journal} {\bibinfo  {journal} {Journal of physics. Condensed matter : an
  Institute of Physics journal}\ }\textbf {\bibinfo {volume} {23}},\ \bibinfo
  {pages} {153102} (\bibinfo {year} {2011})}\BibitemShut {NoStop}%
\bibitem [{\citenamefont {Tka\v{c}ik}\ \emph {et~al.}(2009)\citenamefont
  {Tka\v{c}ik}, \citenamefont {Walczak},\ and\ \citenamefont
  {Bialek}}]{Tkavcik2009}%
  \BibitemOpen
  \bibfield  {author} {\bibinfo {author} {\bibfnamefont {G.}~\bibnamefont
  {Tka\v{c}ik}}, \bibinfo {author} {\bibfnamefont {A.~M.}\ \bibnamefont
  {Walczak}}, \ and\ \bibinfo {author} {\bibfnamefont {W.}~\bibnamefont
  {Bialek}},\ }\href@noop {} {\bibfield  {journal} {\bibinfo  {journal}
  {Physical Review E}\ }\textbf {\bibinfo {volume} {80}},\ \bibinfo {pages}
  {031920} (\bibinfo {year} {2009})}\BibitemShut {NoStop}%
\bibitem [{\citenamefont {Walczak}\ \emph {et~al.}(2010)\citenamefont
  {Walczak}, \citenamefont {Tka\v{c}ik},\ and\ \citenamefont
  {Bialek}}]{Walczak2010}%
  \BibitemOpen
  \bibfield  {author} {\bibinfo {author} {\bibfnamefont {A.~M.}\ \bibnamefont
  {Walczak}}, \bibinfo {author} {\bibfnamefont {G.}~\bibnamefont {Tka\v{c}ik}},
  \ and\ \bibinfo {author} {\bibfnamefont {W.}~\bibnamefont {Bialek}},\ }\href
  {\doibase 10.1103/PhysRevE.81.041905} {\bibfield  {journal} {\bibinfo
  {journal} {Physical Review E}\ }\textbf {\bibinfo {volume} {81}},\ \bibinfo
  {pages} {041905} (\bibinfo {year} {2010})}\BibitemShut {NoStop}%
\bibitem [{\citenamefont {Tka\v{c}ik}\ \emph
  {et~al.}(2012{\natexlab{a}})\citenamefont {Tka\v{c}ik}, \citenamefont
  {Walczak},\ and\ \citenamefont {Bialek}}]{Tkavcik2012}%
  \BibitemOpen
  \bibfield  {author} {\bibinfo {author} {\bibfnamefont {G.}~\bibnamefont
  {Tka\v{c}ik}}, \bibinfo {author} {\bibfnamefont {A.~M.}\ \bibnamefont
  {Walczak}}, \ and\ \bibinfo {author} {\bibfnamefont {W.}~\bibnamefont
  {Bialek}},\ }\href@noop {} {\bibfield  {journal} {\bibinfo  {journal}
  {Physical Review E}\ }\textbf {\bibinfo {volume} {85}},\ \bibinfo {pages}
  {041903} (\bibinfo {year} {2012}{\natexlab{a}})}\BibitemShut {NoStop}%
\bibitem [{\citenamefont {Mugler}\ \emph {et~al.}(2009)\citenamefont {Mugler},
  \citenamefont {Walczak},\ and\ \citenamefont {Wiggins}}]{Mugler2009}%
  \BibitemOpen
  \bibfield  {author} {\bibinfo {author} {\bibfnamefont {A.}~\bibnamefont
  {Mugler}}, \bibinfo {author} {\bibfnamefont {A.}~\bibnamefont {Walczak}}, \
  and\ \bibinfo {author} {\bibfnamefont {C.}~\bibnamefont {Wiggins}},\
  }\href@noop {} {\bibfield  {journal} {\bibinfo  {journal} {Physical Review
  E}\ }\textbf {\bibinfo {volume} {80}},\ \bibinfo {pages} {041921} (\bibinfo
  {year} {2009})}\BibitemShut {NoStop}%
\bibitem [{\citenamefont {Rieckh}\ and\ \citenamefont
  {Tka\v{c}ik}(2014)}]{Rieckh2014}%
  \BibitemOpen
  \bibfield  {author} {\bibinfo {author} {\bibfnamefont {G.}~\bibnamefont
  {Rieckh}}\ and\ \bibinfo {author} {\bibfnamefont {G.}~\bibnamefont
  {Tka\v{c}ik}},\ }\href@noop {} {\bibfield  {journal} {\bibinfo  {journal}
  {Biophysical Journal}\ }\textbf {\bibinfo {volume} {106}},\ \bibinfo {pages}
  {1194} (\bibinfo {year} {2014})}\BibitemShut {NoStop}%
\bibitem [{\citenamefont {Sokolowski}\ and\ \citenamefont
  {Tka\v{c}ik}(2015)}]{Sokolowski2015}%
  \BibitemOpen
  \bibfield  {author} {\bibinfo {author} {\bibfnamefont {T.~R.}\ \bibnamefont
  {Sokolowski}}\ and\ \bibinfo {author} {\bibfnamefont {G.}~\bibnamefont
  {Tka\v{c}ik}},\ }\href@noop {} {\enquote {\bibinfo {title} {Optimizing
  information flow in small genetic networks. iv. spatial coupling},}\ }
  (\bibinfo {year} {2015}),\ \bibinfo {note}
  {http://arxiv.org/abs/1501.04015}\BibitemShut {NoStop}%
\bibitem [{\citenamefont {Tostevin}\ and\ \citenamefont {ten
  Wolde}(2009)}]{Tostevin2009}%
  \BibitemOpen
  \bibfield  {author} {\bibinfo {author} {\bibfnamefont {F.}~\bibnamefont
  {Tostevin}}\ and\ \bibinfo {author} {\bibfnamefont {P.~R.}\ \bibnamefont {ten
  Wolde}},\ }\href {\doibase 10.1103/PhysRevLett.102.218101} {\bibfield
  {journal} {\bibinfo  {journal} {Physical Review Letters}\ }\textbf {\bibinfo
  {volume} {102}},\ \bibinfo {pages} {218101} (\bibinfo {year}
  {2009})}\BibitemShut {NoStop}%
\bibitem [{\citenamefont {Tostevin}\ and\ \citenamefont {ten
  Wolde}(2010)}]{Tostevin2010}%
  \BibitemOpen
  \bibfield  {author} {\bibinfo {author} {\bibfnamefont {F.}~\bibnamefont
  {Tostevin}}\ and\ \bibinfo {author} {\bibfnamefont {P.~R.}\ \bibnamefont {ten
  Wolde}},\ }\href {\doibase 10.1103/PhysRevE.81.061917} {\bibfield  {journal}
  {\bibinfo  {journal} {Physical Review E}\ }\textbf {\bibinfo {volume} {81}},\
  \bibinfo {pages} {061917} (\bibinfo {year} {2010})}\BibitemShut {NoStop}%
\bibitem [{\citenamefont {de~Ronde}\ \emph {et~al.}(2010)\citenamefont
  {de~Ronde}, \citenamefont {Tostevin},\ and\ \citenamefont {ten
  Wolde}}]{Ronde2010}%
  \BibitemOpen
  \bibfield  {author} {\bibinfo {author} {\bibfnamefont {W.~H.}\ \bibnamefont
  {de~Ronde}}, \bibinfo {author} {\bibfnamefont {F.}~\bibnamefont {Tostevin}},
  \ and\ \bibinfo {author} {\bibfnamefont {P.~R.}\ \bibnamefont {ten Wolde}},\
  }\href {\doibase 10.1103/PhysRevE.82.031914} {\bibfield  {journal} {\bibinfo
  {journal} {Physical Review E}\ }\textbf {\bibinfo {volume} {82}},\ \bibinfo
  {pages} {031914} (\bibinfo {year} {2010})}\BibitemShut {NoStop}%
\bibitem [{\citenamefont {de~Ronde}\ \emph {et~al.}(2012)\citenamefont
  {de~Ronde}, \citenamefont {Tostevin},\ and\ \citenamefont {ten
  Wolde}}]{Ronde2012}%
  \BibitemOpen
  \bibfield  {author} {\bibinfo {author} {\bibfnamefont {W.~H.}\ \bibnamefont
  {de~Ronde}}, \bibinfo {author} {\bibfnamefont {F.}~\bibnamefont {Tostevin}},
  \ and\ \bibinfo {author} {\bibfnamefont {P.~R.}\ \bibnamefont {ten Wolde}},\
  }\href {\doibase 10.1103/PhysRevE.86.021913} {\bibfield  {journal} {\bibinfo
  {journal} {Physical Review E}\ }\textbf {\bibinfo {volume} {86}},\ \bibinfo
  {pages} {021913} (\bibinfo {year} {2012})}\BibitemShut {NoStop}%
\bibitem [{\citenamefont {Gregor}\ \emph
  {et~al.}(2007{\natexlab{a}})\citenamefont {Gregor}, \citenamefont
  {Wieschaus}, \citenamefont {McGregor}, \citenamefont {Bialek},\ and\
  \citenamefont {Tank}}]{Gregor2007}%
  \BibitemOpen
  \bibfield  {author} {\bibinfo {author} {\bibfnamefont {T.}~\bibnamefont
  {Gregor}}, \bibinfo {author} {\bibfnamefont {E.~F.}\ \bibnamefont
  {Wieschaus}}, \bibinfo {author} {\bibfnamefont {A.~P.}\ \bibnamefont
  {McGregor}}, \bibinfo {author} {\bibfnamefont {W.}~\bibnamefont {Bialek}}, \
  and\ \bibinfo {author} {\bibfnamefont {D.~W.}\ \bibnamefont {Tank}},\
  }\href@noop {} {\bibfield  {journal} {\bibinfo  {journal} {Cell}\ }\textbf
  {\bibinfo {volume} {130}},\ \bibinfo {pages} {141} (\bibinfo {year}
  {2007}{\natexlab{a}})}\BibitemShut {NoStop}%
\bibitem [{\citenamefont {Gregor}\ \emph
  {et~al.}(2007{\natexlab{b}})\citenamefont {Gregor}, \citenamefont {Tank},
  \citenamefont {Wieschaus},\ and\ \citenamefont {Bialek}}]{Gregor2007a}%
  \BibitemOpen
  \bibfield  {author} {\bibinfo {author} {\bibfnamefont {T.}~\bibnamefont
  {Gregor}}, \bibinfo {author} {\bibfnamefont {D.~W.}\ \bibnamefont {Tank}},
  \bibinfo {author} {\bibfnamefont {E.~F.}\ \bibnamefont {Wieschaus}}, \ and\
  \bibinfo {author} {\bibfnamefont {W.}~\bibnamefont {Bialek}},\ }\href@noop {}
  {\bibfield  {journal} {\bibinfo  {journal} {Cell}\ }\textbf {\bibinfo
  {volume} {130}},\ \bibinfo {pages} {153} (\bibinfo {year}
  {2007}{\natexlab{b}})}\BibitemShut {NoStop}%
\bibitem [{\citenamefont {Dubuis}\ \emph {et~al.}(2013)\citenamefont {Dubuis},
  \citenamefont {Tkacik}, \citenamefont {Wieschaus}, \citenamefont {Gregor},\
  and\ \citenamefont {Bialek}}]{Dubuis2013}%
  \BibitemOpen
  \bibfield  {author} {\bibinfo {author} {\bibfnamefont {J.~O.}\ \bibnamefont
  {Dubuis}}, \bibinfo {author} {\bibfnamefont {G.}~\bibnamefont {Tkacik}},
  \bibinfo {author} {\bibfnamefont {E.~F.}\ \bibnamefont {Wieschaus}}, \bibinfo
  {author} {\bibfnamefont {T.}~\bibnamefont {Gregor}}, \ and\ \bibinfo {author}
  {\bibfnamefont {W.}~\bibnamefont {Bialek}},\ }\href {\doibase
  10.1073/pnas.1315642110} {\bibfield  {journal} {\bibinfo  {journal}
  {Proceedings of the National Academy of Sciences of the United States of
  America}\ }\textbf {\bibinfo {volume} {110}},\ \bibinfo {pages} {16301}
  (\bibinfo {year} {2013})}\BibitemShut {NoStop}%
\bibitem [{\citenamefont {Cheong}\ \emph {et~al.}(2011)\citenamefont {Cheong},
  \citenamefont {Rhee}, \citenamefont {Wang}, \citenamefont {Nemenman},\ and\
  \citenamefont {Levchenko}}]{Cheong2011}%
  \BibitemOpen
  \bibfield  {author} {\bibinfo {author} {\bibfnamefont {R.}~\bibnamefont
  {Cheong}}, \bibinfo {author} {\bibfnamefont {A.}~\bibnamefont {Rhee}},
  \bibinfo {author} {\bibfnamefont {C.~J.}\ \bibnamefont {Wang}}, \bibinfo
  {author} {\bibfnamefont {I.}~\bibnamefont {Nemenman}}, \ and\ \bibinfo
  {author} {\bibfnamefont {A.}~\bibnamefont {Levchenko}},\ }\href@noop {}
  {\bibfield  {journal} {\bibinfo  {journal} {Science}\ }\textbf {\bibinfo
  {volume} {334}},\ \bibinfo {pages} {354} (\bibinfo {year}
  {2011})}\BibitemShut {NoStop}%
\bibitem [{\citenamefont {Pahle}\ \emph {et~al.}(2008)\citenamefont {Pahle},
  \citenamefont {Green}, \citenamefont {Dixon},\ and\ \citenamefont
  {Kummer}}]{Pahle2008}%
  \BibitemOpen
  \bibfield  {author} {\bibinfo {author} {\bibfnamefont {J.}~\bibnamefont
  {Pahle}}, \bibinfo {author} {\bibfnamefont {A.~K.}\ \bibnamefont {Green}},
  \bibinfo {author} {\bibfnamefont {C.~J.}\ \bibnamefont {Dixon}}, \ and\
  \bibinfo {author} {\bibfnamefont {U.}~\bibnamefont {Kummer}},\ }\href@noop {}
  {\bibfield  {journal} {\bibinfo  {journal} {BMC Bioinformatics}\ }\textbf
  {\bibinfo {volume} {9}},\ \bibinfo {pages} {139} (\bibinfo {year}
  {2008})}\BibitemShut {NoStop}%
\bibitem [{\citenamefont {Selimkhanov}\ \emph {et~al.}(2014)\citenamefont
  {Selimkhanov}, \citenamefont {Taylor}, \citenamefont {Yao}, \citenamefont
  {Pilko}, \citenamefont {Albeck}, \citenamefont {Hoffmann}, \citenamefont
  {Tsimring},\ and\ \citenamefont {Wollman}}]{Selimkhanov2014}%
  \BibitemOpen
  \bibfield  {author} {\bibinfo {author} {\bibfnamefont {J.}~\bibnamefont
  {Selimkhanov}}, \bibinfo {author} {\bibfnamefont {B.}~\bibnamefont {Taylor}},
  \bibinfo {author} {\bibfnamefont {J.}~\bibnamefont {Yao}}, \bibinfo {author}
  {\bibfnamefont {A.}~\bibnamefont {Pilko}}, \bibinfo {author} {\bibfnamefont
  {J.}~\bibnamefont {Albeck}}, \bibinfo {author} {\bibfnamefont
  {A.}~\bibnamefont {Hoffmann}}, \bibinfo {author} {\bibfnamefont
  {L.}~\bibnamefont {Tsimring}}, \ and\ \bibinfo {author} {\bibfnamefont
  {R.}~\bibnamefont {Wollman}},\ }\href@noop {} {\bibfield  {journal} {\bibinfo
   {journal} {Science}\ }\textbf {\bibinfo {volume} {346}},\ \bibinfo {pages}
  {1370} (\bibinfo {year} {2014})}\BibitemShut {NoStop}%
\bibitem [{\citenamefont {Mancini}\ \emph {et~al.}(2013)\citenamefont
  {Mancini}, \citenamefont {Wiggins}, \citenamefont {Marsili},\ and\
  \citenamefont {Walczak}}]{Mancini2013}%
  \BibitemOpen
  \bibfield  {author} {\bibinfo {author} {\bibfnamefont {F.}~\bibnamefont
  {Mancini}}, \bibinfo {author} {\bibfnamefont {C.~H.}\ \bibnamefont
  {Wiggins}}, \bibinfo {author} {\bibfnamefont {M.}~\bibnamefont {Marsili}}, \
  and\ \bibinfo {author} {\bibfnamefont {A.~M.}\ \bibnamefont {Walczak}},\
  }\href {\doibase 10.1103/PhysRevE.88.022708} {\bibfield  {journal} {\bibinfo
  {journal} {Physical Review E}\ }\textbf {\bibinfo {volume} {022708}},\
  \bibinfo {pages} {1} (\bibinfo {year} {2013})}\BibitemShut {NoStop}%
\bibitem [{\citenamefont {Mancini}\ \emph {et~al.}(2015)\citenamefont
  {Mancini}, \citenamefont {Marsili},\ and\ \citenamefont
  {Walczak}}]{Mancini2015}%
  \BibitemOpen
  \bibfield  {author} {\bibinfo {author} {\bibfnamefont {F.}~\bibnamefont
  {Mancini}}, \bibinfo {author} {\bibfnamefont {M.}~\bibnamefont {Marsili}}, \
  and\ \bibinfo {author} {\bibfnamefont {A.~M.}\ \bibnamefont {Walczak}},\
  }\href@noop {} {\bibfield  {journal} {\bibinfo  {journal} {Journal of
  Statistical Physics}\ ,\ \bibinfo {pages} {1}} (\bibinfo {year}
  {2015})}\BibitemShut {NoStop}%
\bibitem [{\citenamefont {Kepler}\ and\ \citenamefont
  {Elston}(2001)}]{KeplerElston}%
  \BibitemOpen
  \bibfield  {author} {\bibinfo {author} {\bibfnamefont {T.~B.}\ \bibnamefont
  {Kepler}}\ and\ \bibinfo {author} {\bibfnamefont {T.~C.}\ \bibnamefont
  {Elston}},\ }\href {\doibase 10.1016/S0006-3495(01)75949-8} {\bibfield
  {journal} {\bibinfo  {journal} {Biophysical Journal}\ }\textbf {\bibinfo
  {volume} {81}},\ \bibinfo {pages} {3116} (\bibinfo {year}
  {2001})}\BibitemShut {NoStop}%
\bibitem [{\citenamefont {Raj}\ \emph {et~al.}(2006)\citenamefont {Raj},
  \citenamefont {Peskin}, \citenamefont {Tranchina}, \citenamefont {Vargas},\
  and\ \citenamefont {Tyagi}}]{Raj2005}%
  \BibitemOpen
  \bibfield  {author} {\bibinfo {author} {\bibfnamefont {A.}~\bibnamefont
  {Raj}}, \bibinfo {author} {\bibfnamefont {C.~S.}\ \bibnamefont {Peskin}},
  \bibinfo {author} {\bibfnamefont {D.}~\bibnamefont {Tranchina}}, \bibinfo
  {author} {\bibfnamefont {D.~Y.}\ \bibnamefont {Vargas}}, \ and\ \bibinfo
  {author} {\bibfnamefont {S.}~\bibnamefont {Tyagi}},\ }\href {\doibase
  10.1371/journal.pbio.0040309} {\bibfield  {journal} {\bibinfo  {journal}
  {PLoS biology}\ }\textbf {\bibinfo {volume} {4}} (\bibinfo {year} {2006}),\
  10.1371/journal.pbio.0040309}\BibitemShut {NoStop}%
\bibitem [{\citenamefont {Friedman}\ \emph {et~al.}(2006)\citenamefont
  {Friedman}, \citenamefont {Cai},\ and\ \citenamefont {Xie}}]{Friedman2006a}%
  \BibitemOpen
  \bibfield  {author} {\bibinfo {author} {\bibfnamefont {N.}~\bibnamefont
  {Friedman}}, \bibinfo {author} {\bibfnamefont {L.}~\bibnamefont {Cai}}, \
  and\ \bibinfo {author} {\bibfnamefont {X.}~\bibnamefont {Xie}},\ }\href
  {\doibase 10.1103/PhysRevLett.97.168302} {\bibfield  {journal} {\bibinfo
  {journal} {Physical Review Letters}\ }\textbf {\bibinfo {volume} {97}},\
  \bibinfo {pages} {168302} (\bibinfo {year} {2006})}\BibitemShut {NoStop}%
\bibitem [{\citenamefont {Walczak}\ \emph
  {et~al.}(2005{\natexlab{a}})\citenamefont {Walczak}, \citenamefont {Sasai},\
  and\ \citenamefont {Wolynes}}]{Walczak2005b}%
  \BibitemOpen
  \bibfield  {author} {\bibinfo {author} {\bibfnamefont {A.~M.}\ \bibnamefont
  {Walczak}}, \bibinfo {author} {\bibfnamefont {M.}~\bibnamefont {Sasai}}, \
  and\ \bibinfo {author} {\bibfnamefont {P.~G.}\ \bibnamefont {Wolynes}},\
  }\href {\doibase 10.1529/biophysj.104.050666} {\bibfield  {journal} {\bibinfo
   {journal} {Biophysical journal}\ }\textbf {\bibinfo {volume} {88}},\
  \bibinfo {pages} {828} (\bibinfo {year} {2005}{\natexlab{a}})}\BibitemShut
  {NoStop}%
\bibitem [{\citenamefont {Cai}\ \emph {et~al.}(2006)\citenamefont {Cai},
  \citenamefont {Friedman},\ and\ \citenamefont {Xie}}]{Cai2006}%
  \BibitemOpen
  \bibfield  {author} {\bibinfo {author} {\bibfnamefont {L.}~\bibnamefont
  {Cai}}, \bibinfo {author} {\bibfnamefont {N.}~\bibnamefont {Friedman}}, \
  and\ \bibinfo {author} {\bibfnamefont {X.~S.}\ \bibnamefont {Xie}},\ }\href
  {\doibase 10.1038/nature04599} {\bibfield  {journal} {\bibinfo  {journal}
  {Nature}\ }\textbf {\bibinfo {volume} {440}},\ \bibinfo {pages} {358}
  (\bibinfo {year} {2006})}\BibitemShut {NoStop}%
\bibitem [{\citenamefont {So}\ \emph {et~al.}(2011)\citenamefont {So},
  \citenamefont {Ghosh}, \citenamefont {Zong}, \citenamefont {Sep{\'{u}}lveda},
  \citenamefont {Segev},\ and\ \citenamefont {Golding}}]{Golding2011}%
  \BibitemOpen
  \bibfield  {author} {\bibinfo {author} {\bibfnamefont {L.-h.}\ \bibnamefont
  {So}}, \bibinfo {author} {\bibfnamefont {A.}~\bibnamefont {Ghosh}}, \bibinfo
  {author} {\bibfnamefont {C.}~\bibnamefont {Zong}}, \bibinfo {author}
  {\bibfnamefont {L.~A.}\ \bibnamefont {Sep{\'{u}}lveda}}, \bibinfo {author}
  {\bibfnamefont {R.}~\bibnamefont {Segev}}, \ and\ \bibinfo {author}
  {\bibfnamefont {I.}~\bibnamefont {Golding}},\ }\href {\doibase
  10.1038/ng.821.GENERAL} {\bibfield  {journal} {\bibinfo  {journal} {Nature
  genetics}\ }\textbf {\bibinfo {volume} {43}},\ \bibinfo {pages} {554}
  (\bibinfo {year} {2011})}\BibitemShut {NoStop}%
\bibitem [{\citenamefont {Desponds}\ \emph {et~al.}(2016)\citenamefont
  {Desponds}, \citenamefont {Tran}, \citenamefont {Ferraro}, \citenamefont
  {Lucas}, \citenamefont {Dostatni},\ and\ \citenamefont
  {Walczak}}]{Desponds2016}%
  \BibitemOpen
  \bibfield  {author} {\bibinfo {author} {\bibfnamefont {J.}~\bibnamefont
  {Desponds}}, \bibinfo {author} {\bibfnamefont {H.}~\bibnamefont {Tran}},
  \bibinfo {author} {\bibfnamefont {T.}~\bibnamefont {Ferraro}}, \bibinfo
  {author} {\bibfnamefont {T.}~\bibnamefont {Lucas}}, \bibinfo {author}
  {\bibfnamefont {N.}~\bibnamefont {Dostatni}}, \ and\ \bibinfo {author}
  {\bibfnamefont {A.~M.}\ \bibnamefont {Walczak}},\ }\href {\doibase
  10.1371/journal.pcbi.1005256} {\bibfield  {journal} {\bibinfo  {journal}
  {PLoS computational biology}\ ,\ \bibinfo {pages} {1}} (\bibinfo {year}
  {2016})}\BibitemShut {NoStop}%
\bibitem [{\citenamefont {Cover}\ and\ \citenamefont
  {Thomas}(1991)}]{Cover1991}%
  \BibitemOpen
  \bibfield  {author} {\bibinfo {author} {\bibfnamefont {T.}~\bibnamefont
  {Cover}}\ and\ \bibinfo {author} {\bibfnamefont {J.}~\bibnamefont {Thomas}},\
  }\href@noop {} {\emph {\bibinfo {title} {{Elements of Information Theory}}}}\
  (\bibinfo  {publisher} {John Wiley},\ \bibinfo {address} {New York, New York,
  USA},\ \bibinfo {year} {1991})\BibitemShut {NoStop}%
\bibitem [{\citenamefont {Crooks}(1998)}]{Crooks1998}%
  \BibitemOpen
  \bibfield  {author} {\bibinfo {author} {\bibfnamefont {G.~E.}\ \bibnamefont
  {Crooks}},\ }\href@noop {} {\bibfield  {journal} {\bibinfo  {journal}
  {Journal of Statistical Physics}\ }\textbf {\bibinfo {volume} {90}},\
  \bibinfo {pages} {1481} (\bibinfo {year} {1998})}\BibitemShut {NoStop}%
\bibitem [{\citenamefont {Tome}\ and\ \citenamefont
  {de~Oliveira}(2012)}]{Tome2012}%
  \BibitemOpen
  \bibfield  {author} {\bibinfo {author} {\bibfnamefont {T.}~\bibnamefont
  {Tome}}\ and\ \bibinfo {author} {\bibfnamefont {M.~J.}\ \bibnamefont
  {de~Oliveira}},\ }\href {\doibase 10.1103/PhysRevLett.108.020601} {\bibfield
  {journal} {\bibinfo  {journal} {Physical Review Letters}\ }\textbf {\bibinfo
  {volume} {020601}},\ \bibinfo {pages} {1} (\bibinfo {year}
  {2012})}\BibitemShut {NoStop}%
\bibitem [{\citenamefont {Hornos}\ \emph {et~al.}(2005)\citenamefont {Hornos},
  \citenamefont {Schultz}, \citenamefont {Innocentini}, \citenamefont {Wang},
  \citenamefont {Walczak}, \citenamefont {Onuchic},\ and\ \citenamefont
  {Wolynes}}]{Hornos2005}%
  \BibitemOpen
  \bibfield  {author} {\bibinfo {author} {\bibfnamefont {J.~E.~M.}\
  \bibnamefont {Hornos}}, \bibinfo {author} {\bibfnamefont {D.}~\bibnamefont
  {Schultz}}, \bibinfo {author} {\bibfnamefont {G.~C.~P.}\ \bibnamefont
  {Innocentini}}, \bibinfo {author} {\bibfnamefont {J.}~\bibnamefont {Wang}},
  \bibinfo {author} {\bibfnamefont {A.~M.~W.}\ \bibnamefont {Walczak}},
  \bibinfo {author} {\bibfnamefont {J.~N.}\ \bibnamefont {Onuchic}}, \ and\
  \bibinfo {author} {\bibfnamefont {P.~G.}\ \bibnamefont {Wolynes}},\ }\href
  {\doibase 10.1103/PhysRevE.72.051907} {\bibfield  {journal} {\bibinfo
  {journal} {Physical Review E}\ ,\ \bibinfo {pages} {1}} (\bibinfo {year}
  {2005})}\BibitemShut {NoStop}%
\bibitem [{\citenamefont {Miekisz}\ and\ \citenamefont
  {Szymanska}(2013)}]{Szymanska2011}%
  \BibitemOpen
  \bibfield  {author} {\bibinfo {author} {\bibfnamefont {J.}~\bibnamefont
  {Miekisz}}\ and\ \bibinfo {author} {\bibfnamefont {P.}~\bibnamefont
  {Szymanska}},\ }\href {\doibase 10.1007/s11538-013-9808-7} {\bibfield
  {journal} {\bibinfo  {journal} {Bull Math Biol}\ ,\ \bibinfo {pages} {317}}
  (\bibinfo {year} {2013})}\BibitemShut {NoStop}%
\bibitem [{\citenamefont {Crisanti}\ \emph {et~al.}(2012)\citenamefont
  {Crisanti}, \citenamefont {Puglisi},\ and\ \citenamefont
  {Villamaina}}]{Crisanti2012}%
  \BibitemOpen
  \bibfield  {author} {\bibinfo {author} {\bibfnamefont {A.}~\bibnamefont
  {Crisanti}}, \bibinfo {author} {\bibfnamefont {A.}~\bibnamefont {Puglisi}}, \
  and\ \bibinfo {author} {\bibfnamefont {D.}~\bibnamefont {Villamaina}},\
  }\href {\doibase 10.1103/PhysRevE.85.061127} {\bibfield  {journal} {\bibinfo
  {journal} {Physical Review E}\ }\textbf {\bibinfo {volume} {061127}}
  (\bibinfo {year} {2012}),\ 10.1103/PhysRevE.85.061127}\BibitemShut {NoStop}%
\bibitem [{\citenamefont {Puglisi}\ \emph {et~al.}(2010)\citenamefont
  {Puglisi}, \citenamefont {Pigolotti}, \citenamefont {Rondoni},\ and\
  \citenamefont {Vulpiani}}]{Puglisi2015}%
  \BibitemOpen
  \bibfield  {author} {\bibinfo {author} {\bibfnamefont {A.}~\bibnamefont
  {Puglisi}}, \bibinfo {author} {\bibfnamefont {S.}~\bibnamefont {Pigolotti}},
  \bibinfo {author} {\bibfnamefont {L.}~\bibnamefont {Rondoni}}, \ and\
  \bibinfo {author} {\bibfnamefont {A.}~\bibnamefont {Vulpiani}},\ }\href
  {\doibase 10.1088/1742-5468/2010/05/P05015} {\bibfield  {journal} {\bibinfo
  {journal} {Journal of Statistical Mechanics: Theory and Experiment}\ }\textbf
  {\bibinfo {volume} {05015}} (\bibinfo {year} {2010}),\
  10.1088/1742-5468/2010/05/P05015}\BibitemShut {NoStop}%
\bibitem [{\citenamefont {Busiello}\ \emph {et~al.}()\citenamefont {Busiello},
  \citenamefont {Hidalgo},\ and\ \citenamefont {Maritan}}]{Busiello2019}%
  \BibitemOpen
  \bibfield  {author} {\bibinfo {author} {\bibfnamefont {D.~M.}\ \bibnamefont
  {Busiello}}, \bibinfo {author} {\bibfnamefont {J.}~\bibnamefont {Hidalgo}}, \
  and\ \bibinfo {author} {\bibfnamefont {A.}~\bibnamefont {Maritan}},\
  }\href@noop {} {\bibfield  {journal} {\bibinfo  {journal} {arxiv}\ }}\Eprint
  {http://arxiv.org/abs/arXiv:1810.01833v2} {arXiv:arXiv:1810.01833v2}
  \BibitemShut {NoStop}%
\bibitem [{\citenamefont {Saunders}\ and\ \citenamefont
  {Howard}(2009)}]{saunders2009}%
  \BibitemOpen
  \bibfield  {author} {\bibinfo {author} {\bibfnamefont {T.~E.}\ \bibnamefont
  {Saunders}}\ and\ \bibinfo {author} {\bibfnamefont {M.}~\bibnamefont
  {Howard}},\ }\href {\doibase 10.1103/PhysRevE.80.041902} {\bibfield
  {journal} {\bibinfo  {journal} {Physical Review E}\ }\textbf {\bibinfo
  {volume} {80}},\ \bibinfo {pages} {041902} (\bibinfo {year}
  {2009})}\BibitemShut {NoStop}%
\bibitem [{\citenamefont {Tka\v{c}ik}\ \emph
  {et~al.}(2012{\natexlab{b}})\citenamefont {Tka\v{c}ik}, \citenamefont
  {Walczak},\ and\ \citenamefont {Bialek}}]{Tkacik2012}%
  \BibitemOpen
  \bibfield  {author} {\bibinfo {author} {\bibfnamefont {G.}~\bibnamefont
  {Tka\v{c}ik}}, \bibinfo {author} {\bibfnamefont {A.~M.}\ \bibnamefont
  {Walczak}}, \ and\ \bibinfo {author} {\bibfnamefont {W.}~\bibnamefont
  {Bialek}},\ }\href {\doibase 10.1103/PhysRevE.85.041903} {\bibfield
  {journal} {\bibinfo  {journal} {Physical Review E}\ }\textbf {\bibinfo
  {volume} {85}},\ \bibinfo {pages} {041903} (\bibinfo {year}
  {2012}{\natexlab{b}})}\BibitemShut {NoStop}%
\bibitem [{\citenamefont {Xiong}\ and\ \citenamefont {Jr}(2003)}]{Xiong2003}%
  \BibitemOpen
  \bibfield  {author} {\bibinfo {author} {\bibfnamefont {W.}~\bibnamefont
  {Xiong}}\ and\ \bibinfo {author} {\bibfnamefont {J.~E.~F.}\ \bibnamefont
  {Jr}},\ }\href@noop {} {\bibfield  {journal} {\bibinfo  {journal} {Nature}\
  }\textbf {\bibinfo {volume} {426}},\ \bibinfo {pages} {460} (\bibinfo {year}
  {2003})}\BibitemShut {NoStop}%
\bibitem [{\citenamefont {Tanaka}\ and\ \citenamefont
  {Augustine}(2008)}]{Tanaka2008}%
  \BibitemOpen
  \bibfield  {author} {\bibinfo {author} {\bibfnamefont {K.}~\bibnamefont
  {Tanaka}}\ and\ \bibinfo {author} {\bibfnamefont {G.~J.}\ \bibnamefont
  {Augustine}},\ }\href@noop {} {\bibfield  {journal} {\bibinfo  {journal}
  {Neuron}\ }\textbf {\bibinfo {volume} {59}},\ \bibinfo {pages} {608}
  (\bibinfo {year} {2008})}\BibitemShut {NoStop}%
\bibitem [{\citenamefont {Guisbert}\ \emph {et~al.}(2004)\citenamefont
  {Guisbert}, \citenamefont {Herman}, \citenamefont {Lu},\ and\ \citenamefont
  {Gross}}]{Guisbert2004}%
  \BibitemOpen
  \bibfield  {author} {\bibinfo {author} {\bibfnamefont {E.}~\bibnamefont
  {Guisbert}}, \bibinfo {author} {\bibfnamefont {C.}~\bibnamefont {Herman}},
  \bibinfo {author} {\bibfnamefont {C.~Z.}\ \bibnamefont {Lu}}, \ and\ \bibinfo
  {author} {\bibfnamefont {C.~A.}\ \bibnamefont {Gross}},\ }\href {\doibase
  10.1101/gad.1219204.coli} {\ ,\ \bibinfo {pages} {2812} (\bibinfo {year}
  {2004})}\BibitemShut {NoStop}%
\bibitem [{\citenamefont {Lahav}\ \emph {et~al.}(2004)\citenamefont {Lahav},
  \citenamefont {Rosenfeld}, \citenamefont {Sigal}, \citenamefont
  {Geva-zatorsky}, \citenamefont {Levine}, \citenamefont {Elowitz},\ and\
  \citenamefont {Alon}}]{Lahav2004}%
  \BibitemOpen
  \bibfield  {author} {\bibinfo {author} {\bibfnamefont {G.}~\bibnamefont
  {Lahav}}, \bibinfo {author} {\bibfnamefont {N.}~\bibnamefont {Rosenfeld}},
  \bibinfo {author} {\bibfnamefont {A.}~\bibnamefont {Sigal}}, \bibinfo
  {author} {\bibfnamefont {N.}~\bibnamefont {Geva-zatorsky}}, \bibinfo {author}
  {\bibfnamefont {A.~J.}\ \bibnamefont {Levine}}, \bibinfo {author}
  {\bibfnamefont {M.~B.}\ \bibnamefont {Elowitz}}, \ and\ \bibinfo {author}
  {\bibfnamefont {U.}~\bibnamefont {Alon}},\ }\href {\doibase 10.1038/ng1293}
  {\ \textbf {\bibinfo {volume} {36}},\ \bibinfo {pages} {147} (\bibinfo {year}
  {2004})}\BibitemShut {NoStop}%
\bibitem [{\citenamefont {Tyson}\ and\ \citenamefont
  {Nov{\'{a}}k}(2015)}]{Tyson2015}%
  \BibitemOpen
  \bibfield  {author} {\bibinfo {author} {\bibfnamefont {J.~J.}\ \bibnamefont
  {Tyson}}\ and\ \bibinfo {author} {\bibfnamefont {B.}~\bibnamefont
  {Nov{\'{a}}k}},\ }\href {\doibase 10.1186/s12915-015-0158-9} {\bibfield
  {journal} {\bibinfo  {journal} {BMC Biology}\ ,\ \bibinfo {pages} {1}}
  (\bibinfo {year} {2015})}\BibitemShut {NoStop}%
\bibitem [{\citenamefont {Lucas}\ \emph {et~al.}(2018)\citenamefont {Lucas},
  \citenamefont {Tran}, \citenamefont {{Perez Romero}}, \citenamefont
  {Guillou}, \citenamefont {Fradin}, \citenamefont {Coppey}, \citenamefont
  {Walczak},\ and\ \citenamefont {Dostatni}}]{Lucas2018}%
  \BibitemOpen
  \bibfield  {author} {\bibinfo {author} {\bibfnamefont {T.}~\bibnamefont
  {Lucas}}, \bibinfo {author} {\bibfnamefont {H.}~\bibnamefont {Tran}},
  \bibinfo {author} {\bibfnamefont {C.~A.}\ \bibnamefont {{Perez Romero}}},
  \bibinfo {author} {\bibfnamefont {A.}~\bibnamefont {Guillou}}, \bibinfo
  {author} {\bibfnamefont {C.}~\bibnamefont {Fradin}}, \bibinfo {author}
  {\bibfnamefont {M.}~\bibnamefont {Coppey}}, \bibinfo {author} {\bibfnamefont
  {A.~M.}\ \bibnamefont {Walczak}}, \ and\ \bibinfo {author} {\bibfnamefont
  {N.}~\bibnamefont {Dostatni}},\ }\href@noop {} {\bibfield  {journal}
  {\bibinfo  {journal} {PLoS Genet}\ ,\ \bibinfo {pages} {1}} (\bibinfo {year}
  {2018})}\BibitemShut {NoStop}%
\bibitem [{\citenamefont {Sagawa}\ and\ \citenamefont
  {Ueda}(2012{\natexlab{a}})}]{Ueda2012}%
  \BibitemOpen
  \bibfield  {author} {\bibinfo {author} {\bibfnamefont {T.}~\bibnamefont
  {Sagawa}}\ and\ \bibinfo {author} {\bibfnamefont {M.}~\bibnamefont {Ueda}},\
  }\href {\doibase 10.1103/PhysRevLett.109.180602} {\bibfield  {journal}
  {\bibinfo  {journal} {Physical Review Letters}\ }\textbf {\bibinfo {volume}
  {109}},\ \bibinfo {pages} {1} (\bibinfo {year} {2012}{\natexlab{a}})},\
  \Eprint {http://arxiv.org/abs/1206.2479} {arXiv:1206.2479} \BibitemShut
  {NoStop}%
\bibitem [{\citenamefont {Sagawa}\ and\ \citenamefont
  {Ueda}(2012{\natexlab{b}})}]{Sagawa2012}%
  \BibitemOpen
  \bibfield  {author} {\bibinfo {author} {\bibfnamefont {T.}~\bibnamefont
  {Sagawa}}\ and\ \bibinfo {author} {\bibfnamefont {M.}~\bibnamefont {Ueda}},\
  }\href {\doibase 10.1103/PhysRevE.85.021104} {\bibfield  {journal} {\bibinfo
  {journal} {Physical Review E - Statistical, Nonlinear, and Soft Matter
  Physics}\ }\textbf {\bibinfo {volume} {85}},\ \bibinfo {pages} {1} (\bibinfo
  {year} {2012}{\natexlab{b}})},\ \Eprint {http://arxiv.org/abs/1105.3262}
  {arXiv:1105.3262} \BibitemShut {NoStop}%
\bibitem [{\citenamefont {Raser}\ and\ \citenamefont
  {O'Shea}(2004)}]{Raser2004}%
  \BibitemOpen
  \bibfield  {author} {\bibinfo {author} {\bibfnamefont {J.}~\bibnamefont
  {Raser}}\ and\ \bibinfo {author} {\bibfnamefont {E.}~\bibnamefont {O'Shea}},\
  }\href {http://www.sciencemag.org/content/304/5678/1811.short} {\bibfield
  {journal} {\bibinfo  {journal} {Science}\ }\textbf {\bibinfo {volume}
  {304}},\ \bibinfo {pages} {1811} (\bibinfo {year} {2004})}\BibitemShut
  {NoStop}%
\bibitem [{\citenamefont {Walczak}\ \emph
  {et~al.}(2005{\natexlab{b}})\citenamefont {Walczak}, \citenamefont
  {Onuchic},\ and\ \citenamefont {Wolynes}}]{Walczak2005a}%
  \BibitemOpen
  \bibfield  {author} {\bibinfo {author} {\bibfnamefont {A.~M.}\ \bibnamefont
  {Walczak}}, \bibinfo {author} {\bibfnamefont {J.~N.}\ \bibnamefont
  {Onuchic}}, \ and\ \bibinfo {author} {\bibfnamefont {P.~G.}\ \bibnamefont
  {Wolynes}},\ }\href {\doibase 10.1073/pnas.0509547102} {\bibfield  {journal}
  {\bibinfo  {journal} {Proceedings of the National Academy of Sciences of the
  United States of America}\ }\textbf {\bibinfo {volume} {102}},\ \bibinfo
  {pages} {18926} (\bibinfo {year} {2005}{\natexlab{b}})}\BibitemShut {NoStop}%
\bibitem [{\citenamefont {Puglisi}\ and\ \citenamefont
  {Villamaina}(2009)}]{Puglisi_2009}%
  \BibitemOpen
  \bibfield  {author} {\bibinfo {author} {\bibfnamefont {A.}~\bibnamefont
  {Puglisi}}\ and\ \bibinfo {author} {\bibfnamefont {D.}~\bibnamefont
  {Villamaina}},\ }\href {\doibase 10.1209/0295-5075/88/30004} {\bibfield
  {journal} {\bibinfo  {journal} {{EPL} (Europhysics Letters)}\ }\textbf
  {\bibinfo {volume} {88}},\ \bibinfo {pages} {30004} (\bibinfo {year}
  {2009})}\BibitemShut {NoStop}%
\end{thebibliography}
\end{document}